\documentclass[twocolumn,superscriptaddress]{revtex4-1}

\usepackage{color}
\usepackage{xcolor}
\usepackage[]{graphicx}
\usepackage{latexsym}
\usepackage{natbib}
\usepackage{amssymb}
\usepackage{amsmath}
\usepackage[caption=false]{subfig}
\usepackage{multirow}

\usepackage{dsfont}

\usepackage[final]{feynmp}
\newcommand{\rhoAA}[2]{\rho_{{#1}_A\rightarrow {#2}_A}}
\newcommand{\rhoBB}[2]{\rho_{{#1}_B\rightarrow {#2}_B}}
\newcommand{\phiAB}[2]{\varphi_{{#1}_A\to {#2}_B}}
\newcommand{\phiBA}[2]{\varphi_{{#1}_B\to {#2}_A}}

\newcommand{\drhoAA}[2]{\partial\rho_{{#1}_A\rightarrow {#2}_A}}
\newcommand{\drhoBB}[2]{\partial\rho_{{#1}_B\rightarrow {#2}_B}}
\newcommand{\dphiAB}[2]{\partial\varphi_{{#1}_A\to {#2}_B}}
\newcommand{\dphiBA}[2]{\partial\varphi_{{#1}_B\to {#2}_A}}

\newcommand{\A}[2]{A^{\rm in}_{\,#1_A\,#2_A}}
\newcommand{\AB}[2]{A^{\rm out}_{\,#1_A\,#2_B}}
\newcommand{\B}[2]{A^{\rm in}_{\,#1_B\,#2_B}}
\newcommand{\BA}[2]{A^{\rm out}_{\,#1_B\,#2_A}}

\newcommand{\kinA}[1]{k^{\rm in}_{#1_A}}
\newcommand{\koutA}[1]{k^{\rm out}_{#1_A}}
\newcommand{\kinB}[1]{k^{\rm in}_{#1_B}}
\newcommand{\koutB}[1]{k^{\rm out}_{#1_B}}

\newcommand{\dA}[2]{\delta_{\,#1_A\,#2_A}}
\newcommand{\dB}[2]{\delta_{\,#1_B\,#2_B}}

\newcommand{\s}{\sigma}

\newcommand{\la}{\langle}
\newcommand{\ra}{\rangle}

\newcommand{\beq}{\begin{equation}}
\newcommand{\eeq}{\end{equation}}
\newcommand{\bc}{\begin{cases}}
\newcommand{\ec}{\end{cases}}

\begin{document}

\title{Model of Brain Activation Predicts the Neural Collective Influence Map of the Brain}

\author{Flaviano Morone}
\affiliation{Levich Institute and Physics Department, City College of New York, New York, NY 10031, USA}
\author{Kevin Roth}
\affiliation{Levich Institute and Physics Department, City College of New York, New York, NY 10031, USA}
\affiliation{Theoretical Physics, ETH Z{\"u}rich, 8093 Z{\"u}rich, Switzerland}
\author{Byungjoon Min}
\affiliation{Levich Institute and Physics Department, City College of New York, New York, NY 10031, USA}
\author{H. Eugene Stanley}
\affiliation{Center for Polymer Studies and Physics Department, Boston University, Boston, MA 02215}
\author{Hern\'an A. Makse}
\email[]{hmakse@lev.ccny.cuny.edu}
\affiliation{Levich Institute and Physics Department, City College of New York, New York, NY 10031, USA}



\begin{abstract}
Efficient complex systems have a modular structure, but modularity
does not guarantee robustness, because efficiency also requires  an
ingenious interplay of the interacting modular components.  The human
brain is the elemental paradigm of an efficient robust modular system
interconnected as a network of networks (NoN).  Understanding the
emergence of robustness in such modular architectures from the
interconnections of its parts is a long-standing challenge that has
concerned many scientists. Current models of dependencies in NoN
inspired by the power grid express interactions among modules with
fragile couplings that amplify even small shocks, thus preventing
functionality.  Therefore, we introduce a model of NoN to shape the
pattern of brain activations to form a modular environment that is
robust. The model predicts the map of neural collective influencers
(NCIs) in the brain, through the optimization of the influence of the
minimal set of essential nodes responsible for broadcasting
information to the whole-brain NoN. Our results suggest new
intervention protocols to control brain activity by targeting
influential neural nodes predicted by network theory.
\end{abstract}


\maketitle

Experience reveals that the brain is composed of massively
connected neural elements arranged in modules \cite{caldarelli,newman}
spatially distributed yet highly integrated to form a system of
network of networks (NoN)
\cite{tononi,treisman1996binding,dehaene,corbetta,hans,bullmore,saulo}. These
modules integrate in larger aggregates to ensure a high level of global
communication efficiency within the overall brain network, while
preserving an extraordinary robustness against malfunctioning
\cite{tononi,treisman1996binding,dehaene}.

The question of how these different modules integrate to preserve
robustness and functionality is a central problem in systems science
\cite{tononi,treisman1996binding,dehaene}. The simplest modular model
\cite{newman} would assign the same function to the connections inside
the modules and across the modules.  However, the existence of
modularity gives rise to two types of connections of intrinsically
different nature: the  intermodular links and intramodular links
\cite{corbetta,saulo,shlomo,gallos}. Intramodular links define
modules usually composed of clustered nodes that perform the same
specific function, like for instance, the visual cortex responsible
for processing visual information.
Besides having intralinks, nodes in a given module may have
 intermodular connections to control or modulate the activity of nodes
in other spatially remote modules
\cite{dehaene,corbetta,tononi,saulo,sigman}.

For example, in integrative sensory processing, the  intermodular
links mediate the bottom-up (or stimulus-driven) processes from lower-order 
areas (eg, visual) to higher-order cortical ones, and top-down
(or goal-directed) control from higher levels to lower ones
\cite{dehaene,corbetta,tononi,sigman}.  Indeed, in studies of
attention, the pattern of brain activation indicates that high-level
regions in dorsal parietal and frontal cortex are involved in
controlling low-level visuo-spatial areas forming a system of networks
connected through intermodular control links (dorsal-frontoparietal
NoN) \cite{corbetta,sigman}.  The purpose of this work is to introduce
a minimal model for a robust brain NoN made of such intramodule
connections and intermodular controllers, which, by abstracting away
complexity, will allow us to make falsifiable predictions about the
location of the most influential nodes in the brain NoN. Targeting
these neural collective influencers (NCIs) influencers may help in
designing intervention protocols to control brain activity prescribed
by network theory \cite{stam,heuvel}.

\section*{Results}

\begin{figure}
\centering
\includegraphics[width=\columnwidth]{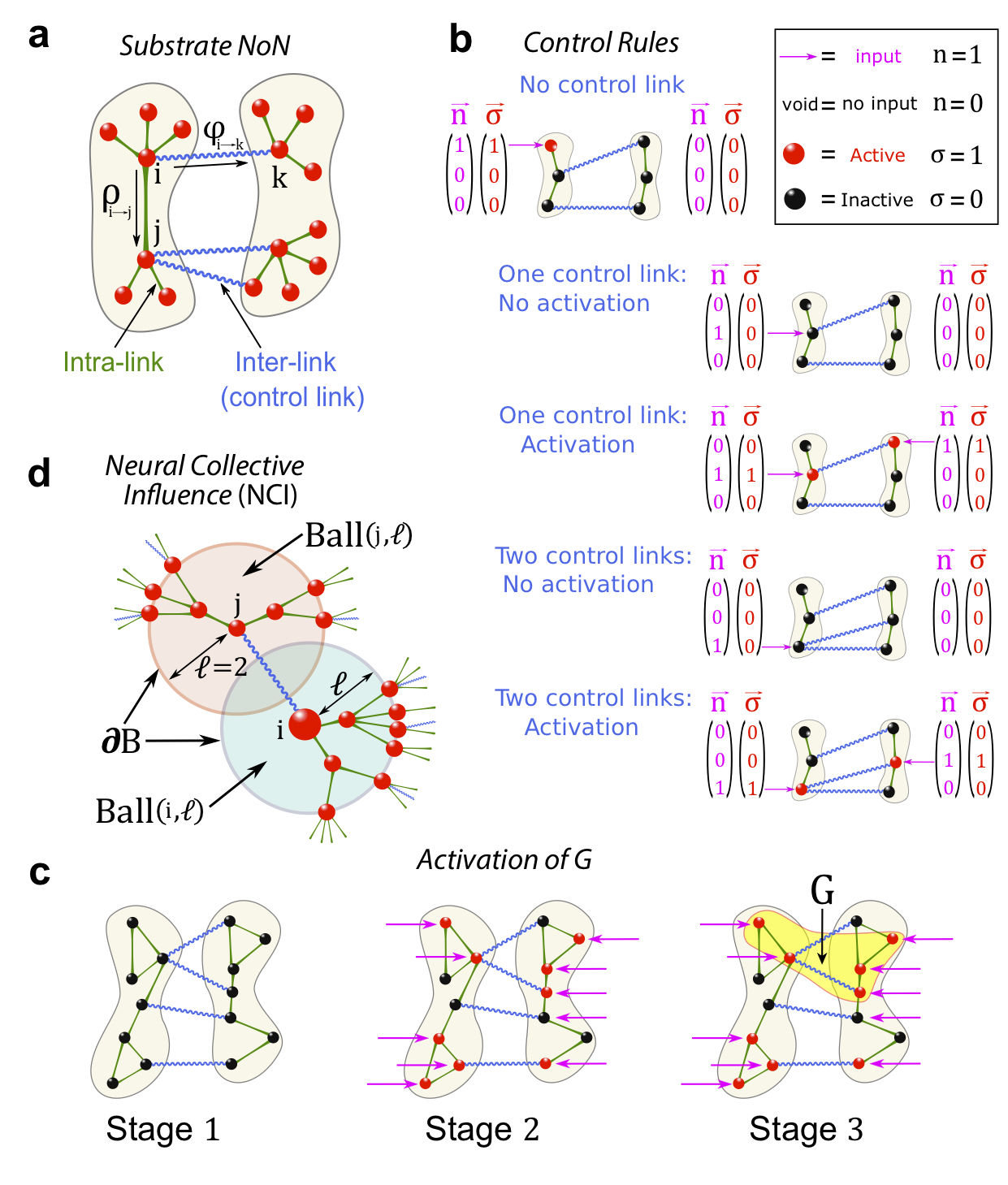}
\caption{
{\bf Definition of NoN model}. {\bf a, Substrate
  NoN.}
Each node has $k_i^{\rm in}$ intramodular links and $k_i^{\rm out}$
intermodular control links. Nodes send information through two
messages to their neighbors: a message $\rho_{i\to j}$ along the
intralink and a message $\varphi_{i\to k}$ along the control link.
{\bf b, Control rule Eq.\,(\ref{eq:sigma2})}. A node $i$ in the
substrate NoN may receive an external input $n_i=1$, or not $n_i=0$.
If the node has no control link it activates as soon as it receives
the external input: $n_i=1=\s_i$.  If it has $1$ control link, it
activates $\s_i=1$ if and only if it receives the input, $n_i=1$, and
its neighbor at the edge of the interlink receives the input as well
($n_j=1$).  If it has $2$ control links (or more) it activates
($\s_i=1$) iff it receives the input and at least one node among its
$j$ neighbors at the edge of the interlink also receives an input
$n_j=1$, otherwise it does not activate ($\s_i=0$).  {\bf c,
  Activation of the mutual giant component}. Global communication in
the NoN is measured through the largest active component $G$ which is
measured only with the active nodes $\s_i=1$. We start with a NoN with
no external input (all $n_i=0$), then $G=0$ (Stage 1). Once an input
is presented to the brain NoN (Stage 2) nodes activate according to
the rules in {\bf b}, and the largest component of activated nodes
defines $G$ (Stage 3), which it is not necessarily equal to the sum of
the individual giant components of the single networks.  Note the
crucial ingredient of the model (not shared by the model of
\cite{shlomo}): active nodes ($\s=1$) may exist outside $G$, and they
can have intermodular control links with other nodes outside the
giant component.  Thus, nodes can be active without being part of the
giant component of their own network in contrast to the rules in
\cite{shlomo}.  {\bf d, Collective Influence}. The collective
influence of node $i$ is determined by the sum of the degree of nodes
in $G$ on the surface of two balls of influence with radius $\ell$:
$\partial {\rm Ball}(i,\ell)$ centered at $i$, and $\partial {\rm
  Ball}(j,\ell)$ centered at $j$, where $j$ is a neighbor of $i$ at
the edge of an interlink having out-degree $k_j^{\rm out}=1$.
}
\label{fig:fig1}
\end{figure}

We consider a substrate NoN composed by two modules
(Fig.\,\ref{fig:fig1}a, below we generalize to more modules).  Every
node $i$ has $k_i^{\rm in}$ intramodular links to nodes in the same
module and $k_i^{\rm out}$ intermodular links to control other
modules (for the sake of simplicity we first consider the case
$k_i^{\rm out}\in\{0, 1\}$ for every node $i$; the general case
$k_i^{\rm out}\in\mathbb{N}_0$ will be treated later).  Because
controlling links connect two different modules, they are fundamentally
different from intramodular ones: the latter encode only the
information about {\it if} two nodes are connected or not inside a
module, whereas the former carry the additional information about {\it
  how} nodes control each other in two
different modules.  We arrive to an important difference between both
types of links which has been recognized in previous NoN models
\cite{shlomo}. An intermodular link between two nodes exists because of
their mutual dependence across two distinct modules performing
different functions.  Therefore, it is reasonable that for this intermodular
link to be active, both nodes across the  
modules should be active.  On
the contrary, nodes inside a module 
connected only via intramodular links that do not
participate in intermodular dependencies will be active independently
on the other module's activity. The intralinks and interlinks are
analogous to the strong and weak links defining hierarchical modules
in the NoN in Refs. \cite{gallos, saulo}.

To elaborate on the mode of intermodular control, think of a node $i$
as a receiver of inputs external to the NoN such as external sensory
inputs to the primary visual cortex (Fig.\,\ref{fig:fig1}b and SI
Text).
The input variable $n_i=1,0$ specifies whether $i$ receives the external
input ($n_i=1$) or not ($n_i=0$).  For example, in the visual system,
$n_i=1$ is the subset of nodes receiving inputs in the earlier stages
in cortical sensory processing \cite{corbetta}.

According to the discussion above, the input $n_i$ alone does not
determine the activation/inactivation state of the node $i$, which we
measure by the state variable $\s_i$ taking values $\s_i=1$ if $i$ is
activated, and $\s_i=0$ if not.  When $i$ has a control link with $j$
in another network, the state $\s_i$ is determined not only by the
input $n_i$, but also by the input $n_j$ to $j$: node $i$ is activated
$\s_i=1$ only when both nodes receive the input ($n_i=1$ and
$n_j=1$). On the contrary, when at least one of the $i, j$ nodes does
not receive input ($n_i=0$ or $n_j=0$), node $i$ is shut down
$\s_i=0$. This top-down and bottom-up control between different
modules is quantified by the following control rule which acts as a
logical AND between two controlling nodes (we consider $k_i^{\rm
  out}=\{0, 1\}$, see Fig.\,\ref{fig:fig1}b):
\begin{equation}
\s_i\ =\ n_i \, \, n_j,\,\,\,\, \qquad \mbox{control rule for
  $k_i^{\rm out}=1$. }
\label{eq:sigma}
\end{equation}

Because not all nodes participate in the control of other nodes, a
certain fraction of them (determined by the degree distribution $P( k^{\rm out})$)
do not establish intermodular links with other nodes, $k_i^{\rm
  out}=0$.  These nodes without control-links (Fig.\,\ref{fig:fig1}b)
activate as soon as they receive an external input, that is
\begin{equation}
\s_i=\ n_i,\,\,\,\, \qquad \mbox{control rule for $k_i^{\rm
    out}=0$.}
\label{eq:sigma1}
\end{equation}

Generalization of the control rule to more than one control link
per node can be done in different ways.  Here, we consider that a node
is activated ($\s_i=1$) iff it receives the input $n_i=1$ and at least
one among the nodes $j$ in another module connected to $i$ via a
control link receives also an input, i.e. $n_j=1$. Otherwise $i$ is
not activated (Fig.\,\ref{fig:fig1}b).  Mathematically:
\beq
\s_i\ =\ n_i\bigg[1-\prod_{j\in\mathcal{F}(i)}(1-n_j)\bigg],\,\,\,\,\,\,
\mbox{general control rule}
\label{eq:sigma2}
\eeq where $\mathcal{F}(i)$ is the set of $k_i^{\rm out}$ nodes
connected to $i$ via intermodular control links. In the following, we
always refer to the general control model Eq.\,(\ref{eq:sigma2}),
unless stated otherwise.

The distinction between $n_i$ and $\s_i$ models the initial sensory
inputs ($n_i$), and the final state response of the brain ($\s_i$) to
those stimuli from top-down and bottom-up influences \cite{corbetta}.
Thus, the final state of the brain network $\s_i$ encodes the brain's
interpretation of the world by modulating external input $n_i$ via
controls Eq.\,(\ref{eq:sigma2}) from other cortical areas
(Fig.\,\ref{fig:fig1}c).  We note that a general model should explain
brain activation even when no external input is applied to the NoN
(e.g. in resting state brain).  This may be accounted for by a
dynamical system that drives the NoN into stable attractors, which in
resting state may no need external input anymore.

Apart from receiving inputs $n_i$ and controlling other nodes via
Eq.\,(\ref{eq:sigma2}), active nodes can also broadcast information to
the network. When all nodes are active, the information sent by a node
can reach the whole brain NoN. If some nodes become inactive,
 i.e. $\s_i=1\to \s_i=0$, the remaining active nodes group together in
 disjoint components of active nodes, such that information starting
 from an active node in one active component cannot reach another
 active node in a different active component.  We quantify the global
 communication efficiency of the brain NoN with the size of the
 largest (giant) mutually-connected active-component $G$ made of
 active nodes $\s_i=1$ (Stage 3 in Figs.\,\ref{fig:fig1}c)
 \cite{shlomo,saulo,gallos}. By strict definition, $G$ could be
 (almost) the entire brain, e.g., a visual stimulus sets off emotional
 cues, memory areas, etc. In what follows, we will restrict the NoN to
 specific systems of interest in the brain, like the visual or motor
 system, which are identifiable by fMRI methods for a particular
 single task.

Each configuration of active/inactive nodes $\vec{\s}= (\s_1,\dots,
\s_N)$ is associated to a specific working mode of the brain.
The plethora of different functions dynamically executed by the brain
\cite{bullmore,treisman1996binding,dehaene,corbetta} results in the
moment-by-moment changes of the configuration $(\s_1,\dots, \s_N)$,
and thus in different values of $G$. The crux of the matter is that,
for typical input configurations $\vec{n}= (n_1,\dots, n_N)$--- i.e.,
the ones produced by the majority of the external (e.g.\ visual)
inputs--- $G$ has to be large enough for a global integration of
information from distributed areas in the brain. In other words, the
brain NoN has to remain globally activated during the acquisition of
different inputs, meaning that $G$ has to be robust, and the more
robust the more states the brain can achieve.
Therefore, a model of a brain NoN must be able to capture such
robustness.

In our statistical mechanics approach, being robust means that the
brain should develop an extensive $G$ for typically sampled
configurations of the external inputs. As a first approximation, we
assume that these inputs are sampled from a flat (random) distribution
of $\vec{n}$. Thus, we first study the robustness of the NoN across
the configurations of states typically sampled by the brain.  The
problem then becomes a classical percolation study of the NoN
\cite{shlomo} following the activation/inactivation rule of
Eq. (\ref{eq:sigma2}).  Having established our model in the normal
brain under typical inputs, we will then move to disease states, which
impede global communication by annihilating focal essential areas in
$G$ \cite{stam,heuvel}.





We calculate $G$ induced by typical random configurations of inputs
$\vec{n}$
as a function of the fraction $q=1-\la n\ra$ of zero inputs (these
zero inputs are analogous to removed nodes in classical percolation
\cite{shlomo,saulo,gallos}) and we show that $G$ remains sizeable even
for high values of $q$, thus probing the robustness of the model NoN.
At a critical value $q_{\rm rand}$, we find the random percolation
critical point $G(q_{\rm rand})=0$ \cite{shlomo,saulo,gallos}
separating a globally connected phase with non-zero $G\left(q<q_{\rm
  rand}\right)>0$ from a disconnected phase $G(q>q_{\rm rand})=0$
composed of fragmented sub-extensive clusters with no giant component
in the thermodynamic limit. The most robust NoN is tantamount to a
system with no disconnected phase, i.e., with a large value of $q_{\rm
  rand}$, ideally $q_{\rm rand}=1$. That is, the brain is robust if it
can sustain a well-defined giant connected component for as many
typical inputs as possible.

The dynamics of information flow in the NoN is defined as
follows. Generally speaking, each node processes activity from
neighboring nodes. Here, we abstract this coding process by
considering that nodes receive information from other nodes via
``messages'' containing the information about their membership in $G$.
Based on the information they receive, nodes broadcast further
messages, until they eventually agree on who belongs to $G$ across the
whole network. Since there are two types of links, we define two types
of messages: $\rho_{i\to j}\in \{0,1\}$ running along an intramodular
link, and $\varphi_{i\to j}\in \{0,1\}$ running along an intermodular
control link, where $\{0,1\}$ represents a $\{$no, yes$\}$ {\it ``I
  belong to $G$''} message, respectively (Fig.\,\ref{fig:fig1}a).

In this view, the existence of an extensive giant mutually-connected
component across the NoN, $G>0$, expresses a percolation phase
produced by the binding of activation patterns across different
modules in a distributed emergent global system. Under this
interpretation, perception is not the responsibility of any particular
cortical area but is an emergent critical property of the percolation
of memberships interchanged across all members of $G$
\cite{crick}. The percolation critical point $q_{\rm rand}$ can be
interpreted as the transition between a phase of global perception
$G>0$ for $q<q_{\rm rand}$ and a null perception phase characterized
by non-extensive disconnected components and the concomitant $G=0$ for
$q>q_{\rm rand}$.

The equations governing the information flow in the brain NoN follow
the updating rules of the membership messages according to (analytical
details in SI Text):
\beq
\begin{aligned}
&\rho_{i\rightarrow j}\ =\ \sigma_{i}
  \Big[\ 1\ - \hspace{-.3cm}\prod_{k \in \mathcal{S}(i) \setminus
      j} \hspace{-.2cm} (1-\rho_{k\rightarrow i} )\hspace{-.1cm}
    \prod_{l \in \mathcal{F}(i)}
\hspace{-.1cm}( 1 - \varphi_{l\to i} )\  \Big]\ ,\\
&\varphi_{i\rightarrow
  j}\ =\ \sigma_{i} \Big[\ 1\ - \hspace{-.3cm}\prod_{k \in
    \mathcal{S}(i)} \hspace{-.2cm} (1-\rho_{k\rightarrow i}
  )\hspace{-.1cm} \prod_{l \in \mathcal{F}(i)\setminus j}
\hspace{-.1cm}( 1 - \varphi_{l\to i} )\ \Big]\ ,
\label{eq:messBNoN}
\end{aligned}
\eeq where $\mathcal{S}(i)\setminus j$ is the set of $k_i^{\rm in}-1$
neighbors of node $i$ in the same module, except $j$.
Equations \,(\ref{eq:messBNoN}) indicate, for instance, that a
positive membership message $\rho_{i\rightarrow j} = 1$ is transmitted
from node $i \to$ node $j$ in the same module (analogously,
$\varphi_{i\rightarrow j}$ transmits messages to the other module) if
node $i$ is active $\s_i=1$ {\it and} if it receives at least one
positive message from either a node $k$ in the same module
$\rho_{k\rightarrow i} = 1$ {\it or} a node $l$ in the other module
$\varphi_{l\rightarrow i} = 1$.  The logical OR is important; it is
the basis for such a robust \mbox{R-NoN} brain model of activation as
elaborated below.


To compute $G$, it is sufficient to know for each node $i$ whether or not it is
a member of $G$, which is encoded in the quantity
$\rho_i\in\{0,1\}$ representing the probability to belong to $G =
\langle \rho_i \rangle$:
\beq \rho_{i}\ =\ \sigma_{i}
\Big[\ 1\ - \hspace{-.3cm}\prod_{k \in \mathcal{S}(i)} \hspace{-.2cm}
  (1-\rho_{k\rightarrow i} )\hspace{-.1cm} \prod_{l \in
    \mathcal{F}(i)}
\hspace{-.1cm}( 1 - \varphi_{l\to i} )\  \Big]\ .
\label{eq:gcomp}
\eeq


Here we arrive to an important point (illustrated in
Fig.\,\ref{fig:fig1}c), which ultimately explains the robustness of
our brain NoN: in our model a node can be active ($\s_i=1$) even if it
does not belong to the giant mutually-connected active component $G$,
thus preventing catastrophic cascading effects. This feature of the
brain model is supported by neuro-anatomical correlates: the brain
responds reasonably well to injuries, for instance, to areas such as
the arcuate fasciculus (the white matter tract that connects the two
most important language areas -- Broca's and Wernicke's area).  This
property is the main difference between our model and previous NoN
models \cite{shlomo} describing catastrophic collapse in power-grids
\cite{rosato}, as discussed next.

{\bf Universality Classes of NoN.--} In the model of
Ref.\,\cite{shlomo}, a node can be active only
if it belongs to the giant component in its own network.  Thus, in
this model the active/inactive state of a node is controlled by the
whole global giant component $\rho_i$, rather than the local state
variable $\s_i$, Eq.\,(\ref{eq:sigma2}), as in our model. This means
that in Ref. \cite{shlomo}, the state of a node is actually controlled
by the whole network [i.e., intermodular controls (therein called
dependencies) carry the weight of the extensive giant component].
Analogously, the NoN cannot be built from the $G=0$ phase, since it
would require the existence of extensive components for each network.
For this reason, the resulting NoN \cite{shlomo} is fragile; a single
inactivation of a node can lead to catastrophic collapse of the whole
active giant component (which, we note, can be avoided by strong
correlations between the hubs of different networks \cite{saulo}).  Conversely, 
the model of Eq.\,(\ref{eq:sigma2}) allows nodes to be
active even if they do not belong to $G$, i. e., when they belong to
non-extensive disconnected components. These small components become
crucial to build the $G>0$ phase from the $G=0$ phase by adding
interlinks to non-extensive components.


Indeed, the model of \cite{shlomo} was proposed to capture the
fragility of certain man-made infrastructures, such as the
catastrophic collapse of power grids--- e.g., the US Northeast
blackout of 2003 which allegedly started in a single power-line
failure as modeled in \cite{shlomo}.  The equation to compute $G$ in
this catastrophic \mbox{C-NoN} model reads: \beq
\rho_{i}\ =\ \sigma_{i} \Big[\ 1\ - \hspace{-.3cm}\prod_{k \in
    \mathcal{S}(i)} \hspace{-.2cm} (1-\rho_{k\rightarrow i}
  )\hspace{.1cm}\Big] \Big[\ 1\ - \hspace{-.2cm} \prod_{k \in
    \mathcal{F}(i)}
\hspace{-.1cm}( 1 - \varphi_{k\to i} )\  \Big]\ .
\label{eq:shlomo}
\eeq We note that Eq.\,(\ref{eq:shlomo}) differs from \mbox{R-NoN}
Eq.\,(\ref{eq:gcomp}) in that the logical OR has been replaced by the
logical AND for message passing in \mbox{C-NoN}.

A third possible model for NoN is the modular model \cite{newman}
mentioned in the introduction which assumes no difference between
intralinks and interlinks as studied in \cite{raissa}. In this model
there are no control-links, therefore, nodes cannot control each
other, and the state equals the input: $\s_i=n_i$. This model is
described using only the intralink messages, $\rho_{i\rightarrow j}$,
corresponding to a single network structure, albeit with modularity
\cite{newman}, and $\rho_{i}$ is simple given by (no special messages
between modules): \beq \rho_{i}\ =\ n_{i}
\Big[\ 1\ - \hspace{-.3cm}\prod_{k \in \mathcal{S}(i)} \hspace{-.2cm}
  (1-\rho_{k\rightarrow i} )\hspace{.1cm}\Big]\ .
\label{eq:raissa}
\eeq 

We thus arrive to three different universality classes of NoN---
\mbox{R-NoN}, \mbox{C-NoN} and modular single network--- according to
the three models given by Eqs.\,(\ref{eq:gcomp}), (\ref{eq:shlomo})
and (\ref{eq:raissa}), respectively, which are defined according to
which variable controls the state of node $i$ ($\s_i$, $\rho_i$,
$n_i$), see Table~\ref{table1}.
Among the three universality classes, only R-NoN is robust with the
functionality of control across modules via top-down and bottom-up
influences.
\begin{table}[b]
\centering
\resizebox{\columnwidth}{!}{
\begin{tabular}{lccc} 
Universality Class & \begin{tabular}{l@{}c@{}}\,\,\,State \\ Control\end{tabular}\,\,\, & \,\,\,Robust\,\,\, & \begin{tabular}{@{}c@{}}Control \\functionality\end{tabular} \\ 
\hline 
\begin{tabular}{l@{}c@{}}Brain Robust \\ \,\,\, R-NoN Eq.\,(\ref{eq:gcomp})\end{tabular} &$\sigma_{i}$ & YES & YES\\ 
\begin{tabular}{l@{}c@{}}Power-Grid Catastrophic\\ \,\,\, C-NoN Eq.\,(\ref{eq:shlomo})\end{tabular} & $\rho_i$ & NO & YES\\ 
\begin{tabular}{l@{}c@{}}Modular Single \\ \,\,\, Network Eq.\,(\ref{eq:raissa}) \end{tabular} & $n_i$ & YES & NO\\
\hline
\end{tabular}
}
\caption{\label{table1}
Universality classes of NoN.}
\end{table}


{\bf Robustness of the brain NoN to typical inputs.--} 
We compute $G(q)$ from Eq.\,(\ref{eq:gcomp}) when we present different
typical random inputs $n_i$ and show that the obtained percolation
threshold $q_{\rm rand}$ is close to 1.
The results are first tested on synthetic NoN made of
Erd\"{o}s-R\'enyi (ER) and scale-free (SF) random graphs
\cite{caldarelli}.

\begin{figure}
\includegraphics[width=\columnwidth]{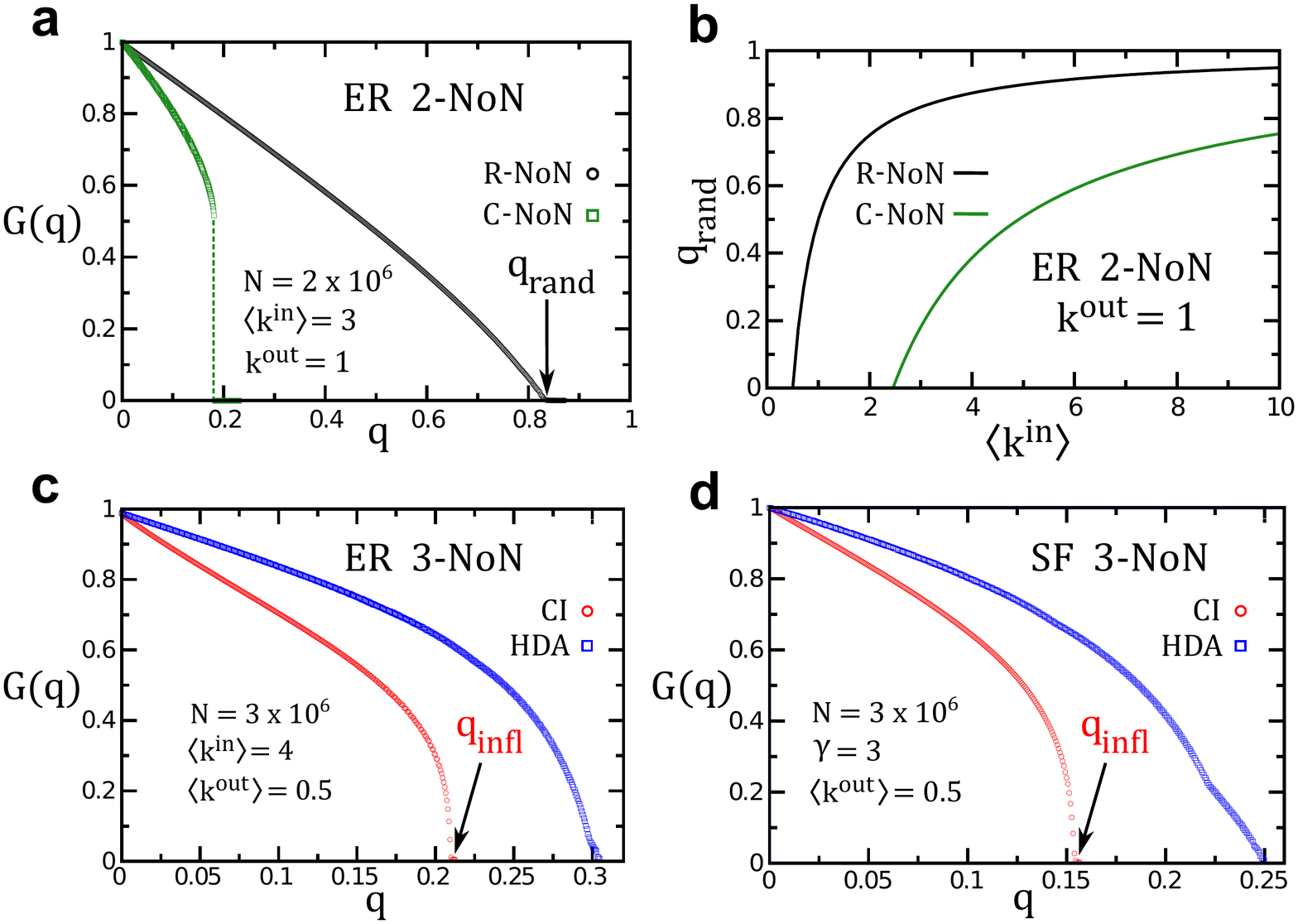}
\caption{ {\bf Robustness and NCI in NoN.}  {\bf a, Robustness of NoN
    under typical random inputs}. Size of the largest active component
  $G(q)$ for typically sampled inputs $\vec{n}$ for ER 2-NoN (meaning
  a NoN made of 2 ER networks) for the \mbox{R-NoN} and \mbox{C-NoN}
  universality classes ($k^{\rm out}=1$ for all nodes, one-to-one
  control links, total size $N= 2\times10^6$). The large value of
  $q_{\rm rand}$ in \mbox{R-NoN} compared to \mbox{C-NoN} confirms the
  robustness of the former. The transition separating the phases $G=0$
  and $G>0$ is $2^{\rm nd}$-order in \mbox{R-NoN} and $1^{\rm
    st}$-order in \mbox{C-NoN}, reinforcing the fundamental difference
  (robust {\it vs} fragile) of these two universality classes (errors
  are s.e.m. over $10$ realizations).  {\bf b, Phase diagram for
    \mbox{R-NoN} and \mbox{C-NoN}}. Behavior of $q_{\rm rand}$ as a
  function of the average $\langle k^{\rm in}\rangle $ for the ER 2-NoN
  in {\bf a}, where each node has $k^{\rm out}=1$. Here $q_{\rm rand}$
  is the fraction of nodes with zero inputs in one network (nodes in
  the other network have all nonzero inputs). The difference in
  $q_{\rm rand}$ between \mbox{R-NoN} and \mbox{C-NoN} ranges from
  $20\%$ for $\langle k^{\rm in}\rangle=10$ to $80\%$ for $\langle
  k^{\rm in}\rangle\sim2.5$.  Analytically, we find for \mbox{R-NoN} with $k^{\rm out} = 1$,
  $q_{\rm rand}=1-1/(2\langle k^{\rm in}\rangle)$.
{\bf c, Rare inputs and NCI in ER 3-NoN}. Size of $G(q)$ as a function
of the untargeted ($n_i=0$) nodes $q$ for a NoN of 3 ER networks
(total size $N=3\times 10^6$). Each network has $10^6$ nodes, $\langle
k^{\rm in}\rangle = 4.0$ and $\langle k^{\rm out} \rangle =0.5$.  We
show the CI optimization (red circles, $\ell=4$) and the high-degree
adaptive (HDA) heuristic (blue squares, removal by highest $k^{\rm
  in}$) \cite{barabasi}. The arrow marks the position of the minimal
fraction of influencers $q_{\rm infl}$, which is smaller than the HDA
centrality (errors are s.e.m. over $10$ realizations). Other heuristic
centralities perform worse than HDA. {\bf d, Rare inputs and NCI in SF 3-NoN}.
$G(q)$ for a NoN with 3 SF networks (total size $N=3\times
10^6$). Each network is SF with $10^6$ nodes, minimum and maximum
degree $ k^{\rm in}_{\rm min} = 2$ and $ k^{\rm in}_{\rm max} = 10^3$,
and power-law exponent $\gamma=3$. The node out-degree is
Poisson-distributed with average $\langle k^{\rm out}\rangle=0.5$
(errors are s.e.m. over $10$ realizations).  The difference between CI
($\ell=3$) and HDA is shown; HDA fails to identify $40\%$ of
influencers.
}
\label{fig:fig2}
\end{figure}

Results in Fig.\,\ref{fig:fig2}a show that our model indeed defines a
robust \mbox{R-NoN} characterized by large $q_{\rm rand}$
Additionally, Fig.\,\ref{fig:fig2}b compares model \mbox{R-NoN}
Eq.\,(\ref{eq:gcomp}) with the catastrophic \mbox{C-NoN} universality
class Eq.\,(\ref{eq:shlomo}) showing that these two models capture two
different phenomena, the former robust with larger $q_{\rm rand}$ and
second-order phase transitions, the latter catastrophic with smaller
$q_{\rm rand}$ with first-order abrupt transitions.

{\bf Response to rare events. Neural Collective Influencers.---}
Having investigated the behavior of the model under typical inputs, we
now study the response of the brain NoN to rare events targeting a set
of neural collective influencers (NCI). These are rare inputs: an
optimal (minimal) set of nodes that when they are shut-down $(n_i=0)$
disintegrates the giant component to $G=0$ employing the smallest
possible fraction of nodes, $q_{\rm infl}$. This is the process of
optimal percolation (rather than classical random percolation treated
above) as defined in \cite{CI}. The malfunction of these neural
influencers could be associated with pathological states of the brain
arising from interruption of global communication in the network
structure such as depression or Alzheimer's disease.  The underlying
conjecture is that these influencers could be responsible for
neurological disorders \cite{stam,heuvel}. At the same time,
activating this minimal set of neural influencers, ($n_i=1$, $\s_i=1$)
would optimally broadcast the information to the entire network
\cite{kempe}. Thus, these neural influencers are also the minimal set
of nodes that provide integration of global activity in the NoN \cite{crick}.

Finding this minimal set is a NP-hard combinatorial optimization
problem \cite{kempe}.  Here, we follow \cite{CI} which developed the
theory of optimal percolation for a system with a single network and
solve the problem in a NoN.  As opposed to random percolation that
identifies $q_{\rm rand}$, optimal percolation identifies the minimal
fraction of influencers $q_{\rm infl}$ that, if removed, optimally
fragment the giant connected component, i.e., with minimal removals
($n_i=0$).  We note that these neural influencers are statistically
rare, i.e., they cannot be obtained by random sampling $\vec{n}$.


 

The mapping to optimal percolation \cite{CI} allows us to find brain
influencers under the approximation of a sparse graph by minimizing
the largest eigenvalue $\lambda(q, {\vec n})$ of a modified
non-backtracking (NB) matrix \cite{hashimoto}
${\mathcal{M}}_{\rho\varphi}\equiv (\partial \rho_{i\rightarrow
  j}/\partial \varphi_{k\rightarrow \ell})_{\rho=\varphi=0}$ of the
NoN over all configurations of inputs ${\vec n}$ having a fraction $q$
of zero inputs (analytical details in SI Text).
The NB matrix $\hat{\mathcal{M}}$
controls the stability of the solution of the broken phase $G=0$.
This solution becomes unstable (i.e., $G$ becomes nonzero) when the
largest eigenvalue is 1.  The minimal set of influencers ${\vec
  n}_{\rm infl}$ and their fraction $q_{\rm infl}$ 
are then found by solving: $\lambda(q_{\rm infl},{\vec n}_{\rm infl})
= \min_{{\vec n}}\lambda(q_{\rm infl},{\vec n}) = 1$.



The eigenvalue $\lambda({\vec n})$ can be efficiently minimized by
progressively removing the input ($n_i=1\to n_i=0$) from the nodes
with the highest Collective Influence index ${\rm CI}_{\ell}(i)$
(detailed derivation in SI Text)
given by ($z_i\ \equiv\ k_i^{\rm in}+k_i^{\rm out}-1$): \beq {\rm
  CI}_{\ell}(i)\ =\ z_i\hspace{-.2cm} \sum_{j\in\partial{\rm
    Ball}(i,\ell)}
\hspace{-.3cm}z_j
\hspace{.3cm} +\hspace{.2cm}  
\sum_{\substack{j\in\mathcal{F}(i)\ :\\
\ k_j^{\rm out}\ = 1}}
\hspace{.15cm}
\ z_j\hspace{-.3cm}
\sum_{m\in\partial{\rm Ball}(j,\ell)}z_m\ .
\label{eq:CIformula}
\eeq

The collective influence ${\rm CI}_{\ell}(i)$ of node $i$ is
determined by two factors (see Fig.\,\ref{fig:fig1}d).  The first one
is a {\it node-centric} contribution, given by the first term in
Eq.\,(\ref{eq:CIformula}), where ${\rm Ball}(i,\ell)$ is the set of
nodes inside a ball of radius $\ell>0$ ($\ell$ is the distance of the
shortest path between two nodes), centered on node $i$, and
$\partial{\rm Ball}(i,\ell)$ its frontier. This ball is grown from the
central node $i$ by following both intralinks and interlinks, and
thus may invade different networks in the NoN. The second factor is a
{\it node-eccentric} contribution, given by the second term in
Eq.\,(\ref{eq:CIformula}), where the sum runs over all nodes $j$
connected to $i$ by an interlink which have out-degree equal to one
$k_j^{\rm out}=1$ (this means that nodes $j$ have no other interlinks
except to node $i$). The contribution of each of these $j$ nodes is
given by growing another ball ${\rm Ball}(j,\ell)$ around them. This
last contribution is absent in the single network case \cite{CI}, and
thus, it is a genuine new feature of the brain NoN.

The NCI are formally defined as the nodes in the minimal set upto
$q_{\rm infl}$.  To identify them, we start with all $n_i=1$ and
$\s_i=1$ and we progressively remove one by one the inputs (setting
$n_i=1\to n_i=0$) to the nodes having the largest ${\rm CI}_{\ell}(i)$
value if they are active $\s_i=1$. At each step the ${\rm
  CI}_{\ell}(i)$ values are recomputed, and the algorithm (described
in detail in SI Text)
stops when $G=0$ where the NCI set is identified.

\pagebreak
We first test our predictions on influencers using synthetically
generated ER-NoN and SF-NoN.  
Figures \,\ref{fig:fig2}c and \ref{fig:fig2}d show the optimality
(smaller $q_{\rm infl}$) of our predicted set of influencers in
comparison with the high-degree centrality \cite{barabasi}, a
heuristic commonly used in graph analysis of pathological brain
networks \cite{heuvel}. The theory allows us to predict the neural
collective influence map (NCI-map) of the brain as explained next.

{\bf Neural Collective Influence map of the NoN.---} We apply
our model to a paradigmatic case of stimulus driven attention
\cite{sigman2008,saulo,gallos}.  The experiment consists of a dual
visual-auditory task performed by $16$ subjects (SI Text).
Each subject receives simultaneously a visual stimulus and an auditory
pitch, to which the subject has to respond with the right hand if a
number was larger than a reference and with the left hand if a tone
was of high frequency.

The rationale to choose this experiment, where stimuli are received
simultaneously, is that this imposes to select an appropriate response
order with consequent deployment of high level control modules in the
brain \cite{sigman2008}. This effect emphasizes the role of top-down
control of intermodular links that is the main effect we are trying
to capture in our model.

\begin{figure}
\includegraphics[width=\columnwidth]{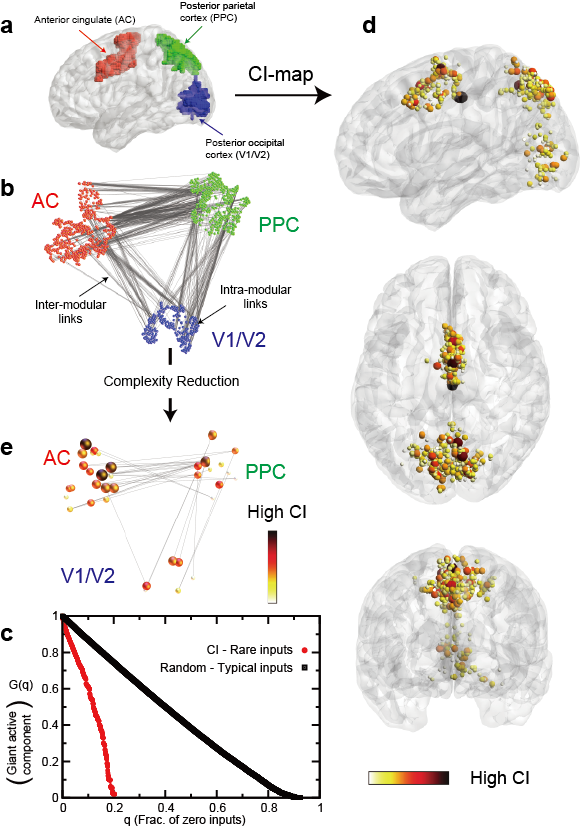}
\caption{ {\bf Brain-NoN.} {\bf a, 3NoN in dual-task fMRI experiment}.
  Spatial location of the 3 main networks for a typical subject (as
  opposed to averaging over all subjects as in {\bf d}) showing the
  anterior cingulate (AC, red), posterior parietal cortex (PPC,
  green), and posterior occipital visual areas V1/V2 (blue). This 3NoN
  structure appears consistently for all 16 subjects.  Nodes in the
  NoN represent voxels in the fMRI BOLD signal of normalized size $2
  \times 2 \times 2$ mm$^3$.  {\bf b, Topology of the 3NoN.} Same as
         {\bf a}, but in the network representation with interlinks
         in gray. Number of nodes in NoN is $N=1,134$, $\langle k^{\rm
           in}\rangle =3.2$, and $\langle k^{\rm out}\rangle=2.5$.
         {\bf c, Robustness and NCI.}  Size of the largest active
         cluster $G(q)$ as a function of the untargeted ($n_i=0$)
         nodes $q$ following CI optimization (red curve, $\ell=3$) and
         following typical random states (black, random percolation).
         {\bf d, NCI-map of the human brain} averaged over 16
         subjects. The color code ranges from 0 to 5.2 and represents
         the number of subjects a node appears in the ranked NCI set
         (see SI Text). High-CI influential regions are located mainly
         in the AC module for processing top-down control, whereas the
         influential nodes are rarely located in the lower-level V1/V2
         region.  The PPC region contains a portion of influential
         nodes closer to AC.  {\bf e, Complexity reduction} to top NCI
         nodes.  Controlling links between different networks are
         mainly mediated by top influencers.}
\label{fig:fig3}
\end{figure}
%

The brain NoN was inferred from the brain activity recorded through
functional magnetic resonance imaging (fMRI).  Nodes in the NoN
represent fMRI voxels whose size is given by the normalized spatial
resolution of the fMRI scan $2 \times 2 \times 2$ mm$^3$.  Pairwise
cross-correlation between the BOLD signals of two nodes represents
only indirect correlations (known as the functional connectivity
network) capturing the weighted sum of all possible direct
interactions between two nodes that could arise from the underlying
unknown structural network and others interactions modulating the
activity of neurons \cite{bullmore}. In order to construct the brain
NoN we infer the strength of these interactions between nodes by using
machine learning maximum-entropy methods
\cite{bialek,sarkar,robinson2}, where we maximize the likelihood of
the interactions between nodes given the observed pattern of fMRI
cross-correlations (full details in SI Text).  The resulting NoN is
shown in Figs.\,\ref{fig:fig3}a and b, which are then used to identify
the NCI in the brain network activated for this particular task.






In all subjects we observe (Fig.\,\ref{fig:fig3}a,\,b): (a) a network
partially covering the anterior cingulate (AC) region, recruited for
decision making and therefore processing top-down and bottom-up
control; (b) a network covering the medial part of the posterior
parietal cortex (PPC) which receives somatosensory inputs and sends
the output to areas of the frontal motor cortex to control particular
movements of the arms; and (c) a network covering the medial part of
the posterior occipital cortex (area V1/V2), along the calcarine
fissure, which is responsible for processing visual information at
lower input levels (an additional auditory network was also observed,
see SI Text).



We apply our theory to the AC-PPC-V1/V2 3NoN to first test the
robustness under typical inputs and then obtain the NCI (rare
inputs). Indeed, the obtained brain 3NoN is very robust to typical
inputs as shown by the large (close to one) $q_{\rm rand}\approx 0.9$
in Fig.\,\ref{fig:fig3}c, black curve.  On the other hand, the theory
is able to localize the minimal set of NCI with $q_{\rm infl}\approx
0.2$, Fig.\,\ref{fig:fig3}c, red curve.  Using these influential nodes
we construct the NCI-map averaging over all subjects.  The emerging
NCI-map averaged over the 16 subjects is portrayed in
Fig.\,\ref{fig:fig3}d (details in SI Text).
We find that the main influence region (high CI) is located mainly in
the AC module as expected, since AC is a central station of top-down
control. The areas of high influence extends also to a portion of the
PPC involved in both top-down and bottom-up control, and it is less
prominent in the V1/V2 areas, which are mostly involved in processing
input information and bottom-up interactions. Therefore, the NCI-map
of the brain suggests that control is deployed from the higher level
module (AC) towards certain strategic locations in the lower ones
(PPC-V1/V2), and these locations can be predicted by network
theory. The complexity reduction obtained by coarse-graining the whole
NoN to the top NCI in Fig.\,\ref{fig:fig3}e highlights the predicted
strategic areas in the brain.

\section*{Discussion}

We present a minimal model of a robust NoN to describe the
integration of brain modules via control interconnections. The key
point of the model is that a node can be active even if it does not
belong to the giant mutually-connected active-component so that
cascades are not fatal.  While our model is expressed {\it in
  abstracto} by logic relations, it is able to make falsifiable
predictions, e.g., the location of the most influential neural nodes
involved in information processing in the brain.
If confirmed experimentally, our results may have applications of
clinical interest, in that they may help to design therapeutic
protocols to handle pathological network conditions and to retune
diseased network dynamics in specific neurological disorders with
interventions targeted to the activity of the influential nodes
predicted by network theory. On the theoretical side, further
extensions of our model are also possible.  For instance, the model
could be enriched by incorporating temporal dependence of brain
activation, which are relevant for the theoretical description of
synaptic plasticity \cite{Min2017}.


{\bf Acknowledgment.} We thank S. Canals, S. Havlin and
  L. Parra for discussions and M. Sigman for providing the data. 
This work was supported by NSF Grants PHY-1305476 and IIS-1515022; 
 NIH-NIBIB Grant 1R01EB022720, NIH-NCI U54CA137788/U54CA132378; 
 and ARL Grant W911NF-09-2-0053 (ARL Network Science CTA).
 The Boston University work was supported by NSF Grants PHY-1505000,
CMMI-1125290, and CHE-1213217, and by DTRA Grant HDTRA1-14-1-0017 
and DOE Contract DE-AC07-05Id14517.





\pagebreak
\clearpage
\newpage

\hspace{1.5cm}
{\LARGE Supplementary Info}
\section*{Message passing in the brain-NoN}
\label{SI::sec:messagePassing}
The classification of connections into intramodule,
\textit{intra-links}, and intermodule, \textit{inter-links},
 together with an introduction of the mathematical
model describing robust brain Network of Networks (NoN) was provided in the
main article. In the present section we further expand the explanation
of the NoN model and the derivation of the message passing equations
describing the information flow in the brain. In the following
sections, we then tackle the problem of finding the most influential
nodes in the brain-NoN with general configuration of intra- and inter-links. 
We conclude with an explanation of the numerical tests and
the construction of the CI-map of the brain.

\medskip

In what follows, we consider two modules A and B, interconnected by
undirected inter-links, where each module is an independent
network made up of $N_A$ respectively $N_B$ nodes connected via
intra-links ($N=N_A+N_B$). The theoretical approach and indeed
the obtained collective influence formula readily carry over to
arbitrary numbers of modules.

Throughout most of the supplementary sections, we adopt the convention to
explicitly show the node's belonging to either module, i.e. every
index $i_A$ representing a node will be accompanied by the network
label to which the node belongs. Moreover, we denote a node's degree
of undirected intra-links by $k^{\rm in}_{i_A}$ and
undirected inter-links degree by $k^{\rm
  out}_{i_A}$. Furthermore, the input variable $n_{i_A} = 1, 0$\,
specifies whether node $i_A$ receives an external input $(n_{i_A} =
1)$ or not $(n_{i_A}=0)$. It is understood that the same terminology
applies equivalently to nodes $i_B$ in module B.

Following the definition of our brain model, we assume that a node
$i_A$ which is connected to one or several nodes from the other module
is activated ($\s_{i_A}=1$) if it receives an input ($n_{i_A}=1$) and
at least one among the nodes $j_B$ connected to it via an inter-link 
also receives an input ($n_{j_B}=1$), as depicted in
Fig.\,1b. In other words, a node with one or several
inter-link dependencies is inactivated when it does not receive the
input $(n_{i_A} = 0)$, or when the last of its neighbors in the other
module ceases to receive an external input. This interaction is
mathematically formalized by the concept of the state
variable\,\,$\s_{i_A}$:
\begin{equation}
\s_{i_A}\ =\ n_{i_A}\Bigg[1-\prod_{{j_B}\in\mathcal{F}({i_A})}(1-n_{j_B})\Bigg]\ , 
\label{SI::eq:superState}
\end{equation}
where $\mathcal{F}({i_A})$ denotes the set of nodes in module B
connected to $i_A$ via an inter link. For the case that node $i_A$ 
has exactly one inter-link to one node $j_B$ in module B, the
above equation reduces to 
\begin{equation}
\s_{i_A} = n_{i_A} n_{j_B}, \,\,\,\,\,\, \mbox{for one-to-one connections.}
\end{equation}
By convention, we also agree to include in the above equation for
$\s_{i_A}$ the case where node $i_A$ does not have any inter-links 
$k^{\rm out}_{i_A}=0$. In this case, we simply equate
\begin{equation}
\s_{i_A} =
n_{i_A},\,\,\,\,\,\, \mbox{for $k^{\rm out}_{i_A}=0$.}
\end{equation}

\begin{center}
\begin{figure}
\centering
\includegraphics[width=\columnwidth]{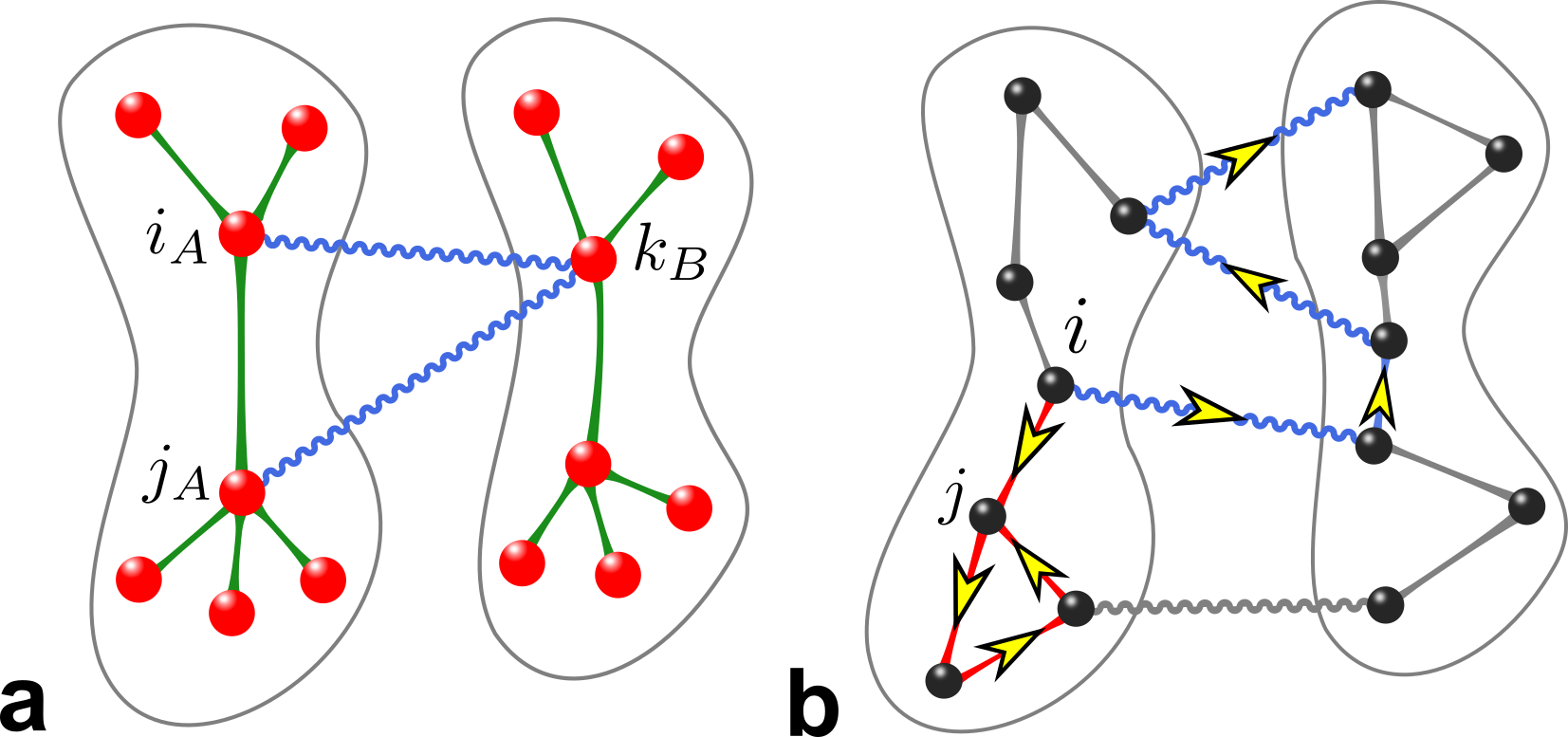}
\caption{
	\label{SI::fig:simpleNoN}
	{\bf a} Simple NoN illustrating the activation rule
  Eq.\,(\ref{SI::eq:superState}). {\bf b} Two NB walks of length $\ell
  = 4$, centered in node $i$ in the 2-NoN. Note that the red walk
  visits node $j$ twice, hence it contains a NB loop. However, as
  shown in \cite{CI}, NB walks with loops can be neglected in the cost
  energy function to leading order $\mathcal{O}(N)$. }
\end{figure}
\end{center}

\noindent Alternatively, we can say that products over empty sets
$\mathcal{F}({i_A})=\emptyset$ default to zero. This is an important
feature of the model, namely that a fraction of nodes determined by $\langle
k^{\rm out} \rangle$ are not involved in control.

In order to get a better understanding of the state variable
$\s_{i_A}$, we consider the following example of the simple NoN
depicted in Fig.\,\ref{SI::fig:simpleNoN}a. 
For this particular
case, we have
\begin{equation}
\begin{aligned}
\s_{i_A}\ & =\ n_{i_A}\,n_{k_B},  \\ \s_{j_A}\ & =\ n_{j_A}\,n_{k_B},
\\ \s_{k_B}\ & =\ n_{k_B} \big[ 1 - ( 1 - n_{i_A} ) ( 1 - n_{j_A} )
  \big],
\end{aligned}
\end{equation}
and the remaining nodes $l$ with no inter-links, $k^{\rm
  out}_l=0$, have $\s_l=n_l$.  

As can be seen, when the nodes in A receive input $n_{i_A} = n_{j_A}=1$
but node $k_B$ does not, $n_{k_B} = 0$, this configuration of external
inputs affects all state variables $\s_{k_B} = \s_{i_A} = \s_{j_A} =
0$. On the other hand, keeping $n_{j_A} = n_{k_B} = 1$ and removing
the input $n_{i_A} = 0$ only affects the state of node $i_A$ by
switching it to inactive $\s_{i_A}=0$ since node $k_B$ is connected to
another node in module A, namely $j_A$, and hence $\s_{k_B} = 1$ is
active together with $\s_{j_A} = 1$.

\medskip
Let us now turn our attention to the messages, representing
information broadcasted between active nodes within the same module or
between active nodes in different modules. The distinction between
intramodule and intermodule messages naturally arises due to the
conceptual difference between intra-links and inter-links
and is reflected in the corresponding distinction between messages
$\rho_{i_A\rightarrow j_A}$ sent along intra-links and messages
$\varphi_{i_A\rightarrow j_B}$ transmitted across inter-links (Fig.\,1a).

It is clear that when all nodes are initially active, the information
is able to circulate in the entire NoN. On the other hand, as
individual nodes are sequentially turned off, the remaining active
nodes are progressively fragmented into disconnected clusters and as a
result the information can no longer be broadcasted globally. The
efficiency to communicate globally can thus be represented by the size
of the \textit{largest (giant) connected cluster of active nodes} $G$ across
all modules constituting the NoN, as 
depicted in Figs.\,1c,\,e.

Formally, we denote 
\begin{equation}
\begin{aligned}
\rhoAA{i}{j} \equiv\ &\textnormal{probability that $i_A$ is connected to $G$}\\
& \textnormal{other than via in-neighbor $j_A$}\ , \\
\phiAB{i}{j} \equiv\ &\textnormal{probability that $i_A$ is connected to $G$}\\ 
& \textnormal{other than via out-neighbor $j_B$}\ .
\end{aligned}
\end{equation}

The size of the mutual giant active component $G$ in turn is entirely determined by the solution of a set of $2M$ self-consistent message passing equations, where $M$ is the total number of intra-links and inter-links in the NoN. 

The derivation of the set of message passing equations corresponding
to our model is provided next.  Let us therefore consider two nodes in
the NoN, say $i_A$ and $j_A$, connected by an intra-link. A node
$i_A$ can send information only if it is active, i.e. if $\s_{i_A} =
1$, and hence the relative message $\rhoAA{i}{j}$ must be proportional
to $\s_{i_A}$. Now, assuming that node $i_A$ is active, it can send a
message to node $j_A$ only if it receives a message by at least one of
its intra-link neighbors other than $j_A$ OR one of its inter-links 
neighbors. Thus, the self-consistent equations describing the
information flow in the brain NoN are given by
\begin{equation}
\begin{aligned}
\rhoAA{i}{j}& = \s_{i_A} \Big[ 1 - \hspace{-.35cm}\prod_{k_A \in
    \mathcal{S}(i_A) \setminus j_A} \hspace{-.5cm} (1-\rhoAA{k}{i}
  )\hspace{-.3cm} \prod_{k_B \in \mathcal{F}(i_A)} \hspace{-.3cm}( 1 -
  \phiBA{k}{i} ) \Big] , 
  \\ \phiAB{i}{j}& = \s_{i_A} \Big[ 1 - \hspace{-.3cm}\prod_{k_A \in
    \mathcal{S}(i_A)} \hspace{-.3cm} (1-\rhoAA{k}{i} )\hspace{-.3cm}
  \prod_{k_B \in \mathcal{F}(i_A)\setminus j_B} \hspace{-.53cm}( 1 -
  \phiBA{k}{i} ) \Big] ,
\end{aligned}
\label{SI::eq:messagePassing}
\end{equation}
where $\mathcal{S}(i_A)$ is the set of intra-link neighbors of node
$i_A$ and $\mathcal{F}(i_A)$ is the set of node $i_A$'s inter-links
neighbors in module B.  
The remaining message passing equations can be obtained by interchanging
the labels for the modules A and B. We note en passant that products
over empty sets $\mathcal{S}(i_A)=\emptyset$ or
$\mathcal{F}(i_A)=\emptyset$ in the above message passing equations
default to one, due to the underlying logical OR in our model.

The size of the mutual giant component $G$ across all modules of the
NoN can then be computed from the fixed point solution for the
intra-link and inter-link messages satisfying the above
self-consistent message passing equations\,(\ref{SI::eq:messagePassing}). 
Explicitly, it is given by
\begin{equation}
G =  \Big( \sum_{i_A = 1}^{N_A} \rho_{i_A} + \sum_{i_B =
  1}^{N_B} \rho_{i_B} \Big) \Big/ ( N_A + N_B)\ ,
\label{SI::eq:giantComp}
\end{equation}
where the 
probability $\rho_{i_A} = 0,1$ for a node $i_A$ to belong to the
largest connected active cluster is computed as
\begin{equation}
\rho_{i_A} = \s_{i_A} \Big[ 1 - \hspace{-.2cm}\prod_{k_A \in
    \mathcal{S}(i_A)} \hspace{-.2cm} (1-\rhoAA{k}{i} ) \hspace{-.2cm}
  \prod_{k_B \in \mathcal{F}(i_A)} \hspace{-.1cm}( 1 - \phiBA{k}{i}
  ) \Big]\ ,
\label{SI::eq:gComp}
\end{equation}
which can be obtained from the expression for the intra-link message
in Eq.\,(\ref{SI::eq:messagePassing}) by including the contribution of
$\rhoAA{j}{i}$ as well.

Strictly speaking, the above message passing equations are valid only
under the assumption that the messages are independent, which is true
for locally tree-like networks, including the thermodynamic limit of
the class of Erd\"{o}s-R\'enyi and scale-free networks
as well as the configuration model (the maximally
random graphs with a given degree distribution \cite{Wormald1981})
which contain loops that grow logarithmically in the system size~\cite{Dorogovtsev2003}. Nevertheless, it is generally accepted, and
confirmed by previous implementations of CI on single networks
\cite{CI}, that results obtained for tree-like graphs apply quite well
also for loopy networks \cite{Altarelli2014,Krzakala2013,Mezard2009}.

\medskip
Next, we turn our attention to two related, but fundamentally
different models. One of them \cite{shlomo}, inspired by the power
grid \cite{rosato}, can be simply obtained from the message passing
equations\,(\ref{SI::eq:messagePassing}) by replacing the underlying
logical OR with a logical AND, as follows
\begin{equation}
\begin{aligned}
\hspace{-5pt}\rho_{i_A\to j_A} &\hspace{-2pt} = \sigma_{i_A}\hspace{-2pt}\Big[ 1 - \hspace{-.5cm}\prod_{k_A
    \in \mathcal{S}(i_A) \setminus j_A} \hspace{-.55cm} (1-\rho_{k_A
    \rightarrow i_A} )\hspace{-1pt}\Big]\hspace{-2pt}\Big[ 1 - \hspace{-.4cm}
  \prod_{k_B \in \mathcal{F}(i_A)} \hspace{-.3cm}( 1 - \varphi_{k_B
    \to i_A} ) \Big] ,\\ 
\hspace{-5pt}\phiAB{i}{j} &\hspace{-2pt}= \s_{i_A}\hspace{-2pt}
\Big[ 1 - \hspace{-.4cm}\prod_{k_A \in
    \mathcal{S}(i_A)} \hspace{-.35cm} (1-\rhoAA{k}{i} )\hspace{-1pt}
  \Big] \hspace{-2pt}\Big[ 1 - \hspace{-.5cm} \prod_{k_B \in
    \mathcal{F}(i_A)\setminus j_B} \hspace{-.5cm}( 1 - \phiBA{k}{i}
  ) \Big] .
\end{aligned}
\label{SI::eq:shlomo_messages}
\end{equation}
In this model, an active node $i_A$ with inter-links to the other
module can send a message $\rhoAA{i}{j}$ to node $j_A$ only if it
receives a message by at least one of its intra-link neighbors other
than $j_A$ AND one of its inter-link neighbors.

Similarly, the probability $\rho_{i_A}$ for a node $i_A$ to belong to
the giant mutually connected active component $G$ can for this model
\cite{shlomo} be obtained by replacing the inherent logical OR in
Eq.\,(\ref{SI::eq:gComp}) with the connective AND:
\begin{equation}
\rho_{i_A} = \sigma_{i_A} \Big[ 1 - \hspace{-.3cm}\prod_{k_A \in
    \mathcal{S}(i_A)} \hspace{-.3cm} (1-\rho_{k_A \rightarrow i_A}
  )\hspace{-0pt}\Big] \hspace{-2pt}\Big[ 1 - \hspace{-.3cm} \prod_{k_B \in
    \mathcal{F}(i_A)} \hspace{-.25cm}( 1 - \varphi_{k_B \to i_A}
  ) \Big] .\\
\label{SI::eq:shlomo_rho}
\end{equation}
We emphasize that Eqs.\,(\ref{SI::eq:shlomo_messages}) and
(\ref{SI::eq:shlomo_rho}) are generalizations of the model
\cite{shlomo}, which considers only one-to-one inter-link (therein called
dependencies), to arbitrary numbers of inter-links.

The third candidate for a NoN to be considered is the simplest possible model, which assumes no difference between intramodule and intermodule connections \cite{newman, raissa} and hence it can be described using only the intra-link messages $\rho_{i\rightarrow j}$, which in this case run along links both within and across modules. Moreover, since there are no dependency links in this model and nodes do not control each other, the state of a node simply equals its input $\s_i = n_i$. The corresponding message passing equations read
\begin{equation}
\begin{aligned}
\rho_{i\rightarrow j} & = n_{i} \Big[ 1 - \hspace{-.3cm}\prod_{k \in \mathcal{S}(i) \setminus j} \hspace{-.2cm} (1-\rho_{k \rightarrow i} )\Big] ,\\
\end{aligned}
\label{SI::eq:structural_messages}
\end{equation}
where for simplicity we dropped the unneeded distinction between different module labels.

The probability $\rho_i$ for a node $i$ to belong to the giant mutually connected active cluster $G$ can again be obtained by taking into account also the contribution from $\rho_{j\rightarrow i}$, as in
\begin{equation}
\begin{aligned}
\rho_{i} & = n_{i} \Big[ 1 - \hspace{-.1cm}\prod_{k \in \mathcal{S}(i)} \hspace{-.05cm} (1-\rho_{k \rightarrow i} )\Big] .\\
\end{aligned}
\label{SI::eq:structural_rho}
\end{equation}

We conclude this discussion by pointing out that the message passing
approach presented in this section not only allows to study
percolation in NoN in a simple and compact way, but it also allows to treat the
non-random removal of inputs and hence investigate the effect of
atypical or rare configurations of inputs on the brain
state. Moreover, the message passing approach allows for an intuitive
interpretation in terms of information flow and can be easily adapted
to include changes in the model as well.

Finally, we recall that the size of the giant mutually connected
active component and indeed the NoN's global communication efficiency
is a function of the input variables $n_{i_A}$ of each node comprising
the NoN. The aim of the next section is thus to find and rank the
minimal set of nodes whose disruption $(n_{i_A} = 1 \rightarrow
n_{i_A} = 0)$ leads to a breakdown of the NoN's global communication
capacity in the most efficient way. We call such nodes
\textit{influencers}.

\section*{Theory of Collective Influence in the brain-NoN}
\label{SI::sec:CItheory}
\subsection*{Derivation of the cost energy function of influence}
\label{SI::subsec:costenergy}

Finding the minimal set of influencers, whose inactivation results in
a breakdown of the NoN's global communication efficiency, is a NP-hard
combinatorial optimization problem originally posed by Kempe {\it et
  al.} \cite{kempe} in the context of maximization of influence in
social network, that is very difficult to solve in general.  In
particular, direct minimization of the size of the mutual giant
component over the configurations of inputs \mbox{$\vec{n} = \{
  n_{1_A}, \dots , n_{N_A}, n_{1_B}, \dots , n_{N_B} \}$} is
untractable, since an explicit functional form of $G(\vec{n})$ is not
feasible.

Instead, the problem of identifying the set of influencers in the
brain NoN can be mapped onto the problem of optimal percolation
\cite{CI}, which, in turn, can be solved by minimizing the largest
eigenvalue $\lambda (\vec{n})$ of the non-backtracking (NB) matrix of
the NoN \cite{CI}. The NB matrix controls the stability of the broken
solution $G=0$ which corresponds to
\mbox{$\{\rhoAA{i}{j}\}=\{\rhoBB{i}{j}\}=\{\phiAB{i}{j}\}=\{\phiBA{i}{j}\}=0$}
and is defined by taking partial derivatives in the message passing
equations\,(\ref{SI::eq:messagePassing}), as follows:
\begin{equation}
\hspace{-2pt}\hat{\mathcal{M}}_{} \equiv \hspace{-2pt}\left. \left(
\begin{array}{cccc}
\hspace{-4pt}\frac{\drhoAA{k}{l}}{\drhoAA{i}{j}} &
\hspace{-5pt}\frac{\drhoBB{k}{l}}{\drhoAA{i}{j}} &
\hspace{-5pt}\frac{\dphiAB{k}{l}}{\drhoAA{i}{j}} &
\hspace{-5pt}\frac{\dphiBA{k}{l}}{\drhoAA{i}{j}} \\[.3cm]
\hspace{-5pt}\frac{\drhoAA{k}{l}}{\drhoBB{i}{j}} &
\hspace{-5pt}\frac{\drhoBB{k}{l}}{\drhoBB{i}{j}} &
\hspace{-5pt}\frac{\dphiAB{k}{l}}{\drhoBB{i}{j}} &
\hspace{-5pt}\frac{\dphiBA{k}{l}}{\drhoBB{i}{j}} \\[.3cm]
\hspace{-5pt}\frac{\drhoAA{k}{l}}{\dphiAB{i}{j}} &
\hspace{-5pt}\frac{\drhoBB{k}{l}}{\dphiAB{i}{j}} &
\hspace{-5pt}\frac{\dphiAB{k}{l}}{\dphiAB{i}{j}} &
\hspace{-5pt}\frac{\dphiBA{k}{l}}{\dphiAB{i}{j}} \\[.3cm]
\hspace{-5pt}\frac{\drhoAA{k}{l}}{\dphiBA{i}{j}} &
\hspace{-5pt}\frac{\drhoBB{k}{l}}{\dphiBA{i}{j}} &
\hspace{-5pt}\frac{\dphiAB{k}{\ell}}{\dphiBA{i}{j}} &
\hspace{-5pt}\frac{\dphiBA{k}{\ell}}{\dphiBA{i}{j}} \\
\end{array} \hspace{-5pt}\right) \right|_{G = 0}\ 
\label{SI::eq:M}
\end{equation}
We note that the NB matrix $\hat{\mathcal{M}}_{i \rightarrow j,\,k
  \rightarrow l}$ is defined over the space of links (see below) and
has non-zero entries only when $(i \rightarrow j,\,k \rightarrow l)$
form a pair of consecutive non-backtracking edges, i.e. $(i
\rightarrow j,\,j \rightarrow l)$ with $i \neq l$ \cite{CI} (see also
Fig.\,\ref{SI::fig:simpleNoN}b). Moreover, powers of the NB matrix count
the number of non-backtracking walks of a given length much in the
same way as powers of adjacency matrices count the number of paths.
 
The minimization of $\lambda (\vec{n})$ is performed over the space of
input configurations\,\,$\vec{n}$ satisfying the condition
$(\sum_{i_A}\hspace{-.04cm}n_{i_A}\hspace{-.08cm}+\sum_{i_B}\hspace{-.03cm}n_{i_B})/(N_A+N_B)
= 1 - q$, where $q$ denotes the fraction of zero inputs. The zero
solution of the message passing equations, corresponding to a
particular configuration $\vec{n}$, is stable if the largest
eigenvalue of the respective NB matrix satisfies $\lambda (\vec{n}) <
1$. Therefore, the optimal configuration $\vec{n}_{\rm infl}$ of
influencers (for which $n_{i_A}, n_{j_B} = 0$), can be found by
solving
\begin{equation}
\lambda(q_{\rm infl}, \vec{n}_{\rm
  infl})\ \equiv\ \min_{{\vec{n}}}\lambda(q_{\rm
  infl},{\vec{n}})\ =\ 1\ ,
\end{equation}
where\, $q_{\rm infl}$\, denotes the minimal fraction of zero inputs,
i.e.\ the influencers. To keep notation light, we shall from now on
omit $q$ in $\lambda(q, \vec{n}) \equiv \lambda (\vec{n})$, which we
assume to be kept fixed.

In order to arrive at an explicit expression for the largest
eigenvalue, we observe that $\lambda (\vec{n})$ determines the growth
rate of an arbitrary non-zero vector $\vec{w}_0$ after $\ell$
iterations with the NB matrix $\hat{\mathcal{M}}$, provided it has
non-vanishing projection onto the corresponding eigenvector. More
precisely, the following equality holds according to the Power Method:
\begin{equation}
\lambda (\vec{n})\ =\ \lim_{\ell \rightarrow \infty} \left[ \frac{\la
    \mathbf{w}_0 |\, \hat{\mathcal{M}}^{\ell}\,| \mathbf{w}_0 \ra}{\la
    \mathbf{w}_0 | \mathbf{w}_0 \ra} \right]^{1/\ell}\ ,
\end{equation}
where $| \mathbf{w}_0 \ra = \vec{w}_0$ denotes the usual column vector
and $\la \mathbf{w}_0 | = \vec{w}_0^{\,\textnormal{T}}$\, denotes the
corresponding row vector.

For finite $\ell$ we define $\la \mathbf{w}_0 |\,
\hat{\mathcal{M}}^{\ell}\,| \mathbf{w}_0 \ra$ to be the cost energy
function of influence at order-$\ell$ and denote the $\ell$-dependent
approximation to the largest eigenvalue
\begin{equation}
\lambda_{\ell} (\vec{n})\ \equiv\ \left[ \frac{\la \mathbf{w}_0\, |
    \hat{\mathcal{M}}^{\ell}\,| \mathbf{w}_0 \ra}{\la \mathbf{w}_0 |
    \mathbf{w}_0 \ra} \right]^{1/\ell}\ .
\end{equation}

In order to derive an analytical expression for $\lambda_{\ell}
(\vec{n})$, it is convenient to elevate the NB matrix
$\hat{\mathcal{M}}$ from the above implicit representation over the
space of
\mbox{$2(M_A\hspace{-.1cm}+\hspace{-.1cm}M_B\hspace{-.1cm}+\hspace{-.1cm}M_{AB})$}
$\times$\,\mbox{$2(M_A\hspace{-.1cm}+\hspace{-.1cm}M_B\hspace{-.1cm}+\hspace{-.1cm}M_{AB})$}
links, where $M_A,\,M_B$ and $M_{AB}$ respectively denote the number of intramodule and intermodule links, and embed it into an enlarged space of dimension
\mbox{$(N_A\hspace{-.1cm}+\hspace{-.1cm}N_B)$}$\times$\mbox{$(N_A\hspace{-.1cm}+\hspace{-.1cm}N_B)$}$\times$\mbox{$(N_A\hspace{-.1cm}+\hspace{-.1cm}N_B)$}$\times$\mbox{$(N_A\hspace{-.1cm}+\hspace{-.1cm}N_B)$}
\cite{CI}.

In this enlarged space, the non-vanishing blocks corresponding to the
NB matrix of our NoN are obtained from Eqs.\,(\ref{SI::eq:messagePassing}) and
are given by (the remaining blocks can be obtained by interchanging
the module labels)
\begin{equation}
\begin{aligned}
\hspace{-3pt}\left.\frac{\drhoAA{k}{l}}{\drhoAA{i}{j}}\right|_{G = 0} & = \s_{k_A} \A{i}{j} \A{k}{l} \dA{j}{k} ( 1 - \dA{i}{l} )  \\
\hspace{-3pt}\left.\frac{\drhoAA{k}{l}}{\dphiBA{i}{j}}\right|_{G = 0} & = \s_{k_A} \BA{i}{j} \A{k}{l} \dA{j}{k}  \\
\hspace{-3pt}\left.\frac{\dphiAB{k}{l}}{\drhoAA{i}{j}}\right|_{G = 0} & = \s_{k_A} \A{i}{j} \AB{k}{l} \dA{j}{k}  \\
\hspace{-3pt}\left.\frac{\dphiAB{k}{l}}{\dphiBA{i}{j}}\right|_{G = 0} & = \s_{k_A} \BA{i}{j} \AB{k}{l} \dA{j}{k} ( 1 - \dB{i}{l} )\ ,
\label{SI::eq:NBblocks}
\end{aligned}
\end{equation}
In the above equations $A$ stands for adjacency matrix and the
superscript '$\rm in$' means that both nodes, represented by the
subscript indices, are within the same module, whereas '$\rm out$'
indicates that they are located in distinct modules. We remind
ourselves that the matrix entries at positions $(i_A, j_B)$ and $(j_B,
i_A)$ are $\AB{i}{j}=\BA{j}{i}=1$ if there exists a connection (in
this case an inter-link) between nodes $i_A$ and $j_B$ and
$\AB{i}{j}=\BA{j}{i}=0$ if there is no connection between these nodes.
The Kronecker deltas reflect the non-backtracking property underlying
the message passing equations\,(\ref{SI::eq:messagePassing}), which
essentially arises due to the fact that a message is computed on the
basis of incoming messages other than from the destination it is sent
to.

Similarly, the intrinsically
\mbox{$2(M_A\hspace{-.1cm}+\hspace{-.1cm}M_B\hspace{-.1cm}+\hspace{-.1cm}M_{AB})$}
dimensional starting vector $\vec{w}_0$, can be embedded into a larger
space of dimension
\mbox{$(N_A\hspace{-.1cm}+\hspace{-.1cm}N_B)$}$\times$\mbox{$(N_A\hspace{-.1cm}+\hspace{-.1cm}N_B)$}.
Without loss of generality, we choose $| \mathbf{w}_0 \ra = |
\mathbf{1} \ra$ as starting vector in the Power Method Iteration,
which translates to $| \mathbf{w}_0 \ra_{i,\,j}\, \equiv\, (\A{i}{j},
\B{i}{j}, \AB{i}{j}, \BA{i}{j})^{\textnormal{T}}$ over the enlarged
vector space.

In what follows, we are going to develop the general \mbox{$\ell$-th} order
expression for the cost energy function of influence corresponding to
the NB matrix of our NoN, which reads

\begin{equation}
\begin{aligned}
\hspace{-6pt}\hat{\mathcal{M}}_{} = \hspace{-1pt}\left.\left(
\begin{array}{cccc}
\label{SI::eq:NBmatrix}
\hspace{-4pt}\frac{\drhoAA{k}{l}}{\drhoAA{i}{j}}\hspace{-4pt} & 0 
& \hspace{-4pt}\frac{\dphiAB{k}{l}}{\drhoAA{i}{j}}\hspace{-4pt} & 0 \hspace{-4pt}\\[.2cm]
0  & \hspace{-4pt}\frac{\drhoBB{k}{l}}{\drhoBB{i}{j}}\hspace{-4pt} 
& 0 & \hspace{-4pt}\frac{\dphiBA{k}{l}}{\drhoBB{i}{j}}\hspace{-4pt}\\[.2cm]
0  & \hspace{-4pt}\frac{\drhoBB{k}{l}}{\dphiAB{i}{j}}\hspace{-4pt} 
& 0 & \hspace{-4pt}\frac{\dphiBA{k}{l}}{\dphiAB{i}{j}}\hspace{-4pt}\\[.2cm]
\hspace{-4pt}\frac{\drhoAA{k}{l}}{\dphiBA{i}{j}}\hspace{-4pt} & 0 
& \hspace{-4pt}\frac{\dphiAB{k}{\ell}}{\dphiBA{i}{j}}\hspace{-4pt} & 0\hspace{-4pt}
\end{array}\hspace{-4pt}\right) \right|_{G = 0}
\end{aligned}
\end{equation}

To this end, we investigate order by order the cost energy function
until the general expression becomes evident. To order $\ell = 1$, we
find
\begin{equation}
\begin{aligned}
& \la \mathbf{w}_0|\, \hat{\mathcal{M}}\, |\mathbf{w}_0\ra\\
& = \sum_{i, j, k, l}^{N_A + N_B}\,{}_{i\,j}\la \mathbf{w}_0 |\,
\hat{\mathcal{M}}_{\,i\,j\,k\,l}\, |\mathbf{w}_0\ra_{k\,l}
\\ & =\ \sum_{i_A}^{N_A}\sum_{j_A}^{N_A} \bigg[
  \sum_{k_A}^{N_A}\sum_{l_A}^{N_A} \A{i}{j} \Big(
  \frac{\drhoAA{k}{l}}{\drhoAA{i}{j}} \Big) \A{k}{l}\hspace{-.1cm}\\ 
  & \hspace{1.5cm}\, + \sum_{k_A}^{N_A}\sum_{l_B}^{N_B} \A{i}{j} \Big(
  \frac{\dphiAB{k}{l}}{\drhoAA{i}{j}} \Big) \AB{k}{l} \bigg] 
\\ & +\ \sum_{i_B}^{N_B}\sum_{j_A}^{N_A} \bigg[
  \sum_{k_A}^{N_A}\sum_{l_A}^{N_A} \BA{i}{j} \Big(
  \frac{\drhoAA{k}{l}}{\dphiBA{i}{j}} \Big) \A{k}{l}\hspace{-.1cm}\\ 
 & \hspace{1.5cm}\, + \sum_{k_A}^{N_A}\sum_{l_B}^{N_B} \BA{i}{j} \Big(
  \frac{\dphiAB{k}{l}}{\dphiBA{i}{j}} \Big) \AB{k}{l} \bigg] \\ 
 & +\ \{ A \leftrightarrow B \}\ ,
\end{aligned}
\end{equation}
where $\{ A \leftrightarrow B \}$ means ``the same terms as above but
with interchanged module labels''.

Inserting the relations for the partial derivatives given by Eq.\,(\ref{SI::eq:NBblocks}) and summing over all independent indices, we obtain the following expression for the cost energy function to lowest order,
\begin{equation}
\begin{aligned}
& \la \mathbf{w}_0|\, \hat{\mathcal{M}}\, |\mathbf{w}_0\ra\\
& =\, \sum_{k_A} \s_{k_A} (\kinA{k} + \koutA{k} \hspace{-1pt}- \hspace{-1pt}1 ) \kinA{k} + \s_{k_A} (\kinA{k} + \koutA{k}\hspace{-1pt} - \hspace{-1pt}1) \koutA{k}\\
& +\, \sum_{k_B} \s_{k_B} ( \kinB{k} + \koutB{k} \hspace{-1pt}- \hspace{-1pt}1) \kinB{k} + \s_{k_B} (\kinB{k} + \koutB{k} \hspace{-1pt}- \hspace{-1pt}1) \koutB{k} .
\end{aligned}
\end{equation}

At this point, it is worth introducing the following notation, which will appear frequently in subsequent expressions for higher order terms
\begin{align}
z_{i_A}\, \equiv\, (\, \kinA{i} + \koutA{i} - 1\, )\ .
\end{align}

This allows us to rewrite even more compactly the final expression for the cost energy function of influence at order $\ell = 1$,
\begin{equation}
\begin{aligned}
& \la \mathbf{w}_0|\, \hat{\mathcal{M}}\, |\mathbf{w}_0\ra\\
 & = \sum_{k_A} \s_{k_A} z_{k_A} (\kinA{k} + \koutA{k})
 + \sum_{k_B} \s_{k_B} z_{k_B} (\kinB{k} + \koutB{k})\, .
\label{SI::eq:cef1}
\end{aligned}
\end{equation}

We proceed to compute the cost energy function to second order from the square of our NB matrix as follows
\begin{equation}
\begin{aligned}
\la \mathbf{w}_0|\, \hat{\mathcal{M}}^2\, |\mathbf{w}_0\ra
 = \sum_{i, j, k, l, m, n}^{N_A + N_B}\hspace{-5pt} {}_{i\,j}\la \mathbf{w}_0| \hat{\mathcal{M}}_{\,i\,j\,k\,l} \hat{\mathcal{M}}_{\,k\,l\,m\,n} |\mathbf{w}_0\ra_{m\,n}  \\
\end{aligned}
\end{equation}
where the matrix elements are given by
\begin{equation}
\begin{aligned}
&  {}_{i\,j}\la \mathbf{w}_0| \hat{\mathcal{M}}_{\,i\,j\,k\,l} \hat{\mathcal{M}}_{\,k\,l\,m\,n} |\mathbf{w}_0\ra_{m\,n}  \\
&\hspace{-.15cm}\begin{array}{rcc}
  = A^{\rm in}_{i_A j_A}\hspace{-2pt}\big( \frac{\drhoAA{k}{l}}{\drhoAA{i}{j}} \big) &\hspace{-.45cm}\Big[ \hspace{-2pt}\big( \frac{\drhoAA{m}{n}}{\drhoAA{k}{l}} \big) A^{\rm in}_{m_A n_A}\hspace{-1pt} + &\hspace{-.35cm}\big( \frac{\dphiAB{m}{n}}{\drhoAA{k}{l}} \big) A^{\rm out}_{m_A n_B} \Big]   \\[.35cm]
 + A^{\rm in}_{i_A j_A}\hspace{-2pt}\big( \frac{\dphiAB{k}{l}}{\drhoAA{i}{j}} \big) &\hspace{-.45cm}\Big[ \hspace{-2pt}\big( \frac{\drhoBB{m}{n}}{\dphiAB{k}{l}} \big) A^{\rm in}_{m_B n_B}\hspace{-1pt} + &\hspace{-.35cm}\big( \frac{\dphiBA{m}{n}}{\dphiAB{k}{l}} \big) A^{\rm out}_{m_B n_A} \Big]   \\[.35cm]
 + A^{\rm out}_{i_B j_A}\hspace{-2pt}\big( \frac{\drhoAA{k}{l}}{\dphiBA{i}{j}} \big) &\hspace{-.45cm}\Big[ \hspace{-2pt}\big( \frac{\drhoAA{m}{n}}{\drhoAA{k}{l}} \big) A^{\rm in}_{m_A n_A}\hspace{-1pt} + &\hspace{-.35cm}\big( \frac{\dphiAB{m}{n}}{\drhoAA{k}{l}} \big) A^{\rm out}_{m_A n_B} \Big]   \\[.35cm]
 +  A^{\rm out}_{i_B j_A}\hspace{-2pt}\big( \frac{\dphiAB{k}{l}}{\dphiBA{i}{j}} \big) &\hspace{-.45cm}\Big[ \hspace{-2pt}\big( \frac{\drhoBB{m}{n}}{\dphiAB{k}{l}} \big) A^{\rm in}_{m_B n_B}\hspace{-1pt} + &\hspace{-.35cm}\big( \frac{\dphiBA{m}{n}}{\dphiAB{k}{l}} \big) A^{\rm out}_{m_B n_A} \Big]   \\[.2cm]
\end{array}\\
&\hspace{-1pt} + \{ A \leftrightarrow B \}  ,
\end{aligned}
\end{equation}

Inserting the appropriate expressions in Eq.\,(\ref{SI::eq:NBblocks}) and summing independent indices, we arrive at
\begin{equation}
\begin{aligned}
& \la \mathbf{w}_0|\, \hat{\mathcal{M}}^2\, |\mathbf{w}_0\ra\\ 
& =\ \sum_{k_A} \s_{k_A} z_{k_A} \big[ \sum_{l_A} \A{k}{l} \s_{l_A} z_{l_A} + \sum_{l_B} \AB{k}{l} \s_{l_B} z_{l_B} \big]  \\
& +\ \sum_{k_B} \s_{k_B} z_{k_B} \big[ \sum_{l_B} \B{k}{l} \s_{l_B} z_{l_B} + \sum_{l_A} \BA{k}{l} \s_{l_A} z_{l_A} \big]\ . 
\end{aligned}
\label{SI::eq:cef2}
\end{equation}

Comparing Eq.\,(\ref{SI::eq:cef1}) for the first order term with Eq.\,(\ref{SI::eq:cef2}) for the second order term, we observe that instead of the in-degree $\kinA{k}$ in the first order expression, we have a sum and the corresponding adjacency matrix $\A{k}{l}$ (multiplied by the factors $\s_{l_A}\,z_{l_A}$) in the second order relation, which together represent exactly $\kinA{k}$\, NB ``steps'' from $k_A$ towards one of the neighboring nodes $l_A \in \mathcal{S}(k_A)$. The generalization of this pattern is of course precisely the NB walk (Fig.\,\ref{SI::fig:simpleNoN}b) in the CI algorithm we are going to derive.

Performing the same analysis as for the previous orders, we find for the cost energy function at order $\ell = 3$,
\begin{equation}
\begin{aligned}
& \la \mathbf{w}_0|\, \hat{\mathcal{M}}^3\, |\mathbf{w}_0\ra\\ 
& = \sum_{k_A} \s_{k_A} z_{k_A} \sum_{l_A} \A{k}{l} \Big[ \sum_{m_A} \A{l}{m} ( 1 \hspace{-1pt}- \hspace{-1pt}\dA{k}{m} ) \s_{m_A} z_{m_A}\\ 
& \hspace{3.45cm}+ \sum_{m_B} \AB{l}{m} \s_{m_B} z_{m_B} \Big]  \\
& + \sum_{k_A} \s_{k_A} z_{k_A} \sum_{l_B} \AB{k}{l} \Big[ \sum_{m_B} \B{l}{m} \s_{m_B} z_{m_B}\\
& \hspace{3.45cm}+ \sum_{m_A} \BA{l}{m} ( 1 \hspace{-1pt}- \hspace{-1pt}\dA{k}{m} ) \s_{m_A} z_{m_A} \Big]  \\
 & +\, \{ A \leftrightarrow B \}\ ,
\label{SI::eq:cef3}
\end{aligned}
\end{equation}
where the factors $( 1 - \dA{k}{m} )$ precisely capture the non-backtracking property of the walks contributing to the cost energy of a given configuration $\vec{n}$, in that they guarantee that the walk never returns to same node it immediately came from.

In general, when we go to higher orders $\ell \geq 4$ of the cost energy function, the NB walk may cross the same node twice and hence contain a NB loop (Fig.\,\ref{SI::fig:simpleNoN}b). It is for instance possible that a NB walk of length 3, which occurs in the cost energy function of influence at order $\ell = 4$, starts and ends in the same node. However, as shown in \cite{CI}, on locally tree-like networks and for large system sizes $N=N_A+N_B$, all NB walks with loops can be neglected to leading order $\mathcal{O} (N)$. 

Therefore, taking into account only the leading order contributions to the cost energy function of influence, we can finally write down the general expression for order $\ell>1$,  
\begin{equation}
\begin{aligned}
\hspace{-6pt}\la \mathbf{w}_0|\, \hat{\mathcal{M}}^{\ell}\, |\mathbf{w}_0\ra & = \sum_{i_A}^{N_A} z_{i_A} \hspace{-3pt}\sum_{j \in \partial {\rm Ball}(i_A, \ell - 1)} \hspace{-2pt}\Big( \prod_{k \in \mathcal{P}_{\ell - 1}(i_A, j)} \hspace{-5pt}\s_k \Big) z_j   \\
& + \sum_{i_B}^{N_B} z_{i_B} \hspace{-3pt}\sum_{j \in \partial {\rm Ball}(i_B, \ell - 1)} \hspace{-2pt}\Big( \prod_{k \in \mathcal{P}_{\ell - 1}(i_B, j)} \hspace{-5pt}\s_k \Big) z_j ,
\label{SI::eq:cef}
\end{aligned}
\end{equation}
where ${\rm Ball}(\,i_A,\,\ell\,)$ is the set of nodes inside a ball
of radius $\ell$ around node $i_A$ (Fig.\,1d), with the radius defined as taking
the shortest path, $\partial{\rm Ball}(\,i_A,\,\ell\,)$ is the
frontier of the ball and $\mathcal{P}_{\ell}(\,i_A,\,j\,)$ is the set
of nodes belonging to the shortest path of length $\ell$ connecting
$i_A$ and $j$. Note that in the above expression the nodes $j$ on the
boundary of the ball as well as the nodes $k$ visited during the
shortest NB walk connecting $i_A$ and $j$ could be in either of the
two modules, which is why we did not explicitly show their module
label.  The corresponding expression for the cost energy function to
order $\ell=1$ is given in Eq.\,(\ref{SI::eq:cef1}).

If we agree to also consider the center node's module label as
implicit, we can write the leading order approximation of the cost
energy function of influence for an arbitrary number of modules to
order $\ell > 1$ as:
\begin{align}
\la \mathbf{w}_0|\, \hat{\mathcal{M}}^{\ell}\, |\mathbf{w}_0\ra & = \sum_{i}\, z_{i} \hspace{-3pt}\sum_{j \in \partial {\rm Ball}(i, \ell - 1)} \hspace{-3pt}\Big( \prod_{k \in \mathcal{P}_{\ell - 1}(i, j)} \s_k \Big) z_j\, .
\label{SI::eq:implicit_cef}
\end{align}

The lowest order expression for arbitrary numbers of modules is given by
\begin{align}
\la \mathbf{w}_0|\, \hat{\mathcal{M}}\, |\mathbf{w}_0\ra & = \sum_{i}\, \s_{i}\, z_{i}\,  (\, k^{\rm in}_i + k^{\rm out}_i\, )\ .\label{SI::eq:implicit_cef1}
\end{align}

As stated in the beginning of this section, the problem of identifying
the optimal set of influencers can be solved by minimizing the largest
eigenvalue $\lambda(\vec{n})$ of the NB matrix corresponding to the
NoN, which we related to the minimization of the leading order
approximation of the the cost energy function of influence given by
Eqs.\,(\ref{SI::eq:implicit_cef}) and (\ref{SI::eq:implicit_cef1}). In
what follows, we propose an efficient algorithm to find the minimal
set of influencers.

\subsection*{Collective Influence algorithm for NoN, CI-NoN}
\label{SI::subsec:CIalgorithm}

Having shown that the minimal set of influencers, whose removal of
input causes a breakdown of the giant mutually connected active
component $G$, can be found by minimizing the cost energy function of
influence, we now proceed to derive the actual minimization protocol,
which we call the Collective Influence algorithm.

Among all the nodes receiving an input, we want to know which node
$i_A$ or $i_B$ in either of the two modules causes the largest drop in
the cost energy function of influence when its input is removed
$(n_{i_A}=1 \rightarrow n_{i_A}=0)$ or \mbox{$(n_{i_B}=1 \rightarrow
n_{i_B}=0)$}.

Let us therefore briefly review the example of the simple NoN depicted
in Fig.\,\ref{SI::fig:simpleNoN} and answer this question for the cost
energy function to order $\ell = 1$, as given in
Eq.\,(\ref{SI::eq:cef1}), assuming that all nodes initially receive an
input.  The important observation to be made here is that removing the
input to node $k_B$, i.e. setting ($n_{k_B}=1 \rightarrow n_{k_B} =
0$) affects all three state variables $\s_{k_B}=\s_{i_A}=\s_{j_A}=0$
and hence decreases the cost energy function by the contribution from
all of the three inactivated nodes, whereas removing the input to
either node $i_A$ or node $j_A$ only affects their own contribution to
the cost energy function. A moment's thought reveals that the crucial
characteristic of node $k_B$, leading to such a deactivation pattern,
is that both of its neighbors $i_A$ and $j_A$ have exactly one
inter-link to $k_B$, i.e. their intermodule degree is
precisely $k^{\rm out}_{i_A} = 1$ and $k^{\rm out}_{j_A} = 1$. In this
case, node $k_B$'s input is pivotal to the activation/deactivation of
its inter-link neighbors $i_A$ and $j_A$.

If we formally define $\textnormal{CI}^{\rm centric}_{\,\ell}(i_A)$ to
be the contribution to the cost energy function of influence at order
$\ell + 1$ centered in $i_A$ and proportional to $\s_{i_A}$,
then $i_A$'s Collective Influence $\textnormal{CI}_{\,\ell}\,(i_A)$ is
the sum of its own $\textnormal{CI}^{\rm centric}_{\,\ell}(i_A)$ and
the $\textnormal{CI}^{\rm centric}_{\,\ell}(j_B)$ of all nodes $j_B$
in the other module with exactly one inter-link to $i_A$
(Fig.\,1d).  We call the sum of the $\textnormal{CI}^{\rm
  centric}_{\,\ell}(j_B)$ of all nodes $j_B$ with $\koutB{j}=1$ the
\textit{eccentric} contribution $\textnormal{CI}^{\rm
  eccentric}_{\,\ell}(i_A)$ to node $i_A$'s Collective Influence.

For an arbitrary number of modules, we define the Collective Influence
of node $i$ as
\begin{equation}
\begin{aligned}
\textnormal{CI}_{l = 0}( i ) & = z_{i}\,( k^{\rm in}_i + k^{\rm
  out}_i )\, + \sum_{\substack{\,j\in\mathcal{F}(i)\, :\\[.05cm]
    \ k_j^{\rm out}\, =\, 1\ }} \hspace{-2pt}z_{j}\, ( k^{\rm in}_j + k^{\rm
  out}_j )\ ,\\[.2cm] \textnormal{CI}_{l \geq 1} ( i) &
= z_{i} \hspace{-.2cm}\sum_{j \in \partial {\rm Ball} (i,
  \ell )} \hspace{-.2cm}  z_{j}\, +
\sum_{\substack{\,j\in\mathcal{F}(i)\, :\\[.05cm] \ k_j^{\rm out}\,
    =\, 1\ }} \,z_{j} \hspace{-.2cm}\sum_{m \in \partial \rm Ball
  (j, \ell )} \hspace{-.2cm}  z_{m}\ ,
\end{aligned}
\label{SI::eq:CIformula}
\end{equation}
where $z_i\, \equiv\, k^{\rm in}_i + k^{\rm out}_i - 1$.  Here ${\rm
  Ball}(\hspace{0.3mm}i,\ell\hspace{0.3mm})$ is the set of nodes
inside a ball of radius $\ell$ centered around node $i$
(Fig.\,1d), with the radius defined as taking the
shortest path and $\partial{\rm Ball}(\hspace{0.3mm}i,
\ell\hspace{0.3mm})$ denotes the set of nodes residing on the frontier
of the ball. We emphasize that nodes on the boundary of the ball can
be in either of the modules.  Indeed, the ball is grown from the
central node following both intra and inter-links and thus
may invade different modules of the brain NoN.  Finally, we remark
that the \textit{node-eccentric} contribution to node $i$'s Collective
Influence, given by the second term in Eq.\,(\ref{SI::eq:CIformula}), is
absent in the single network case \cite{CI} and thus presents a
genuine new feature of the brain NoN.

With the Collective Influence measure~(\ref{SI::eq:CIformula}) at our
disposal, we now proceed to specify the algorithmic implementation to
find and rank the minimal set of influencers ensuring global
communication in the brain NoN.

The \textbf{Collective Influence algorithm} is defined as follows:
Starting from the fully activated NoN, where every node is receiving
an input $n_i = 1$, we progressively remove one by one the inputs
$(n_i = 1 \rightarrow n_i = 0)$ corresponding to the node which has
the largest ${\rm CI}_{\ell} (i)$ value~(\ref{SI::eq:CIformula}),
provided it is active $\s_i = 1$ (Fig.\,1e). After every
removal of an input, the degrees of the removed node's neighbors are updated and the ${\rm CI_{\ell}}$ values of the remaining
active nodes are recomputed from where a new top-CI is removed and so
on.  The algorithm terminates when the largest active mutually
connected component $G$ is zero. The algorithm's performance increases
by using larger values of the radius $\ell$ of the ${\rm
  Ball}(\hspace{0.3mm}i,\ell\hspace{0.3mm})$, which must however not
exceed the original diameter of the NoN, for otherwise the Collective
Influence is zero ${\rm CI}_{\ell} (i) = 0$.  In practice, we observe
that already for $\ell = 3, 4$ the algorithm reaches the top
performance (Figs.\,2c,\,d).

The Collective Influence theory developed above allows us to compute
the minimal fraction $q_{\rm infl}$ as well as the actual
configuration $\vec{n}_{\rm infl}$ of influencers whose removal
annihilates the giant active component $G$ and therefore brings the
NoN's global communication efficiency to a halt. In the case $q <
q_{\rm infl}$, however, the giant component is nonzero, a consequence
of the fact that the system of Eqs.\,(\ref{SI::eq:messagePassing}) has
another stable solution different from
\mbox{$\{\rhoAA{i}{j}\}=\{\rhoBB{i}{j}\}=\{\phiAB{i}{j}\}=\{\phiBA{i}{j}\}$}
identically zero: $G=0$. Therefore, for $q < q_{\rm infl}$ the
stability of the new solution $G(q) \neq 0$ is not controlled by
the NB operator anymore, but a more complicated operator comes into play that
depends on the form of the solution itself.  The solution to this
problem was presented in \cite{CI} and consists in implementing a
reinsertion scheme. The reinsertion rule used to obtain the ${\rm CI}$
curves shown in Figs.\,2c,\,d follows the one presented
in \cite{CI} and is defined as follows: given the minimal set of
influencers up to $q_{\rm infl}$, we reinsert one by one the inputs
$(n_i = 0 \rightarrow n_i = 1)$ corresponding to the node $i$ which
joins the smallest number of active clusters in the NoN when
reinserted $n_i = 1$. In practice, we reinserted a finite fraction of
the total number of inputs that were removed to break the giant
component, before recomputing again the number of clusters the
influencers to be reinserted would join. We arrive in this way to the
minimal set of influencers ranked from top CI to zero. This list is
then used to rank the nodes in the brain.

\section*{Method to construct the brain NoN}
\label{si-brain}

\subsection*{Dual task experiment}
\label{si-dual}

Our brain networks rely on functional magnetic resonance imaging
(fMRI).  The fMRI data consists of time-series of the blood oxygen
level dependent (BOLD) signals based on phase and amplitude response
to a dual task involving visual and auditory stimuli obtained for each
voxel.  We use the dual-task experiment on humans explained in detail
in Refs.\,\cite{sigman2007,sigman2008,saulo,gallos}.  The data that we
used in this study can be found at:
\url{http://www-levich.engr.ccny.cuny.edu/webpage/hmakse/software-and-data}.
The experiment is part of a larger neuroimaging research program headed by
Denis Le Bihan and approved by the Comit\'e Consultatif pour la
Protection des Personnes dans la Recherche Biom\'edicale, H\^opital de
Bic\^etre (Le Kremlin-Bic\^etre, France).

Sixteen participants (7 women and 9 men, mean age, 23, ranging from 20
to 28) performed a dual-task paradigm: a visual task of comparing an
Arabic number to a fixed reference and an auditory task of judging the
pitch of auditory tone.  The two stimuli were applied to subjects
simultaneously.  Subjects were asked to press a key using right and
left hand, respectively, when the number appearing on the screen was
larger than a reference and the tone was high frequency.

\subsection*{Details of NoN reconstruction}

%

  The fMRI data we used to construct the brain NoN are taken from Ref.\,\cite{sigman2008}.  
  As outlined in great detail in Ref.\,\cite{sigman2008}, a $3$T fMRI detector (Bruker)
  was utilized to record the
  blood oxygenation level-dependent (BOLD) signals from a T$2^*$-weighted
  gradient echoplanar imaging sequence [repetition time
    ${\rm(TR)}=1.5\ s$; echo time = 40 ms; angle = 90$^o$; field of
    view (FOV) = 192 $\times$ 256 mm; matrix 64 $\times$ 64]. 
  Within this setup, the entire brain was obtained in 24 slices with a thickness of 5mm each. 
  The experimenters also recorded high-resolution images (three-dimensional gradient echo
  inversion-recovery sequence, inversion time = 700 mm; FOV = 192
  $\times$ 256 $\times$ 256 mm; matrix = 256 $\times$ 128 $\times$
  256; slice thickness 1 mm).  

  Data analysis in Ref.\,\cite{sigman2008}, was performed with SPM2 software. 
  In order to quantify the phase and periodicity of the fMRI data, 
  the authors of \cite{sigman2008}, regressed the BOLD signal for each participant and trial (8 TRs of
  1.5 s) against a sine and a cosine.
%
%
%
%
%
 %
 %
  To avoid numerical
  instabilities, Ref.\,\cite{sigman2008} detrended the raw signal for each voxel within each trial,
  correcting for linear drifts and subtracting the mean
  (the average phase within
each participant and condition was computed using the appropriate mean for circular quantities). 
The projections of the sine and cosine for each voxel $j$, are given by: \beq A^jx =
  \sum_is_i\cos\left(\frac{2\pi\cdot{\rm TR}\cdot i}{\rm ITI}\right),
  \eeq and \beq A^jy = \sum_is_i\sin\left(\frac{2\pi\cdot{\rm TR}\cdot
    i}{\rm ITI}\right), \eeq where $\{s_i\}$ corresponds to the detrended
  signal, and $j$ denotes the voxel number.  The inter-trial interval ITI was 12 sec, and TR 1.5 sec. 
  To account for anatomical differences in brain morphology when averaging across the participants, Ref.\,\cite{sigman2008}
  stereotactically transformed to the standardized coordinate space of
  Talairach and Tournoux [(Montreal Neurological Institute) MNI 152 average brain] and
  smoothed the regression parameters of the sine and cosine (7 mm full-width at
  half-maximum).
  As described in
  \cite{sigman2007}, phase and amplitude were calculated as \beq
\begin{aligned}
\phi^j\ &= \ {\rm arctan}(A^jy/A^jx)\ ,\\ 
A^j\ &= \sqrt{ (A^jx)^2 + (A^jy)^2}\ , 
\end{aligned}
\label{eq:phase_ampl}
\eeq where $A^jx$ and $A^jy$ denote the regression weights
of the cosine and sine for voxel $j$ respectively. 
The phase was additionally multiplied by $12/2\pi$\,s, 
in order to obtain a fraction of the
stimulation period of $12$ s, with a phase of $0$\,s indicating a peak
activation coinciding with stimulus onset \cite{sigman2008}. 

In order to confine the brain network reconstruction to voxels
  participating in the task setup, \cite{sigman2008} 
  estimated the fraction of the measured phases that are within 
  the expected response range (ERR). 
  Overall, $64$ phase
  measurements, corresponding to four conditions per participant, were obtained. 
  On the basis of previous characterizations
  of the hemodynamic response function, Ref.\,\cite{sigman2008} set the ERR to
  the interval from $2$ to $10$ s, thus allowing for region-to-region and inter-condition variations.
  The probability for a given number of $x$ measurements (out of the 64 total) to lie
  within the ERR can accordingly be calculated from the binomial distribution, as outlined in
  \cite{sigman2007}.  Reference\,\cite{sigman2008} restricted the network analysis
  to voxels with more than $48$ measurements within the ERR, corresponding to
  a binomial probability $p < 0.05$.
It is worth to note that the authors of \cite{sigman2008} evaluated the significance
  of the phase variations with delay using a second-level SPM model which
  contained all the single-trial phase measurements. 

Ref.\,\cite{sigman2008}, performed the following two statistical tests with the collected data.
  First, they searched for linearly increasing phases as a function of delay
  (contrast $-2 -1\ 1\ 2$, accounting for irregularities in the delay spacing). 
  Second, they looked for regions with a delay by
  regime type interaction (contrast $1-1 -1 1 $), corresponding to a ``psychological refractory period''
  PRP effect. Moreover, measurements of the
  single-trial response amplitude were tested with the same SPM model. 

\subsection*{Definition of Brain-NoN}
\label{si-definition}




The construction of the 3NoN composed of AC-PPC-V1/V2 
depicted in Fig.\,3a consists of two main steps: first we identify 
the nodes belonging to each module, and then we create the intra-links 
and the inter-links between them (we remark that intra-links and inter-links are
analogous to the strong links and weak links defined in refs \cite{saulo} and \cite{gallos}). 
In the former step, we use the 
cross-correlation $C_{ij}$ between the phases of BOLD response for
each pair of voxels $i$ and $j$, while in the latter step we use 
a machine learning algorithm to infer the pairwise interactions $J_{ij}$ 
between voxels from the correlations $C_{ij}$. By thresholding the values of 
the $J_{ij}$ we then create the connections between the voxels inside and 
across the modules. In the following discussion, we first explain how 
to identify the nodes in the three modules, and then we move to 
explain how we infer the connections between the nodes. 

\begin{center}
\begin{figure}[]
\centering
\includegraphics[width=0.8\columnwidth]{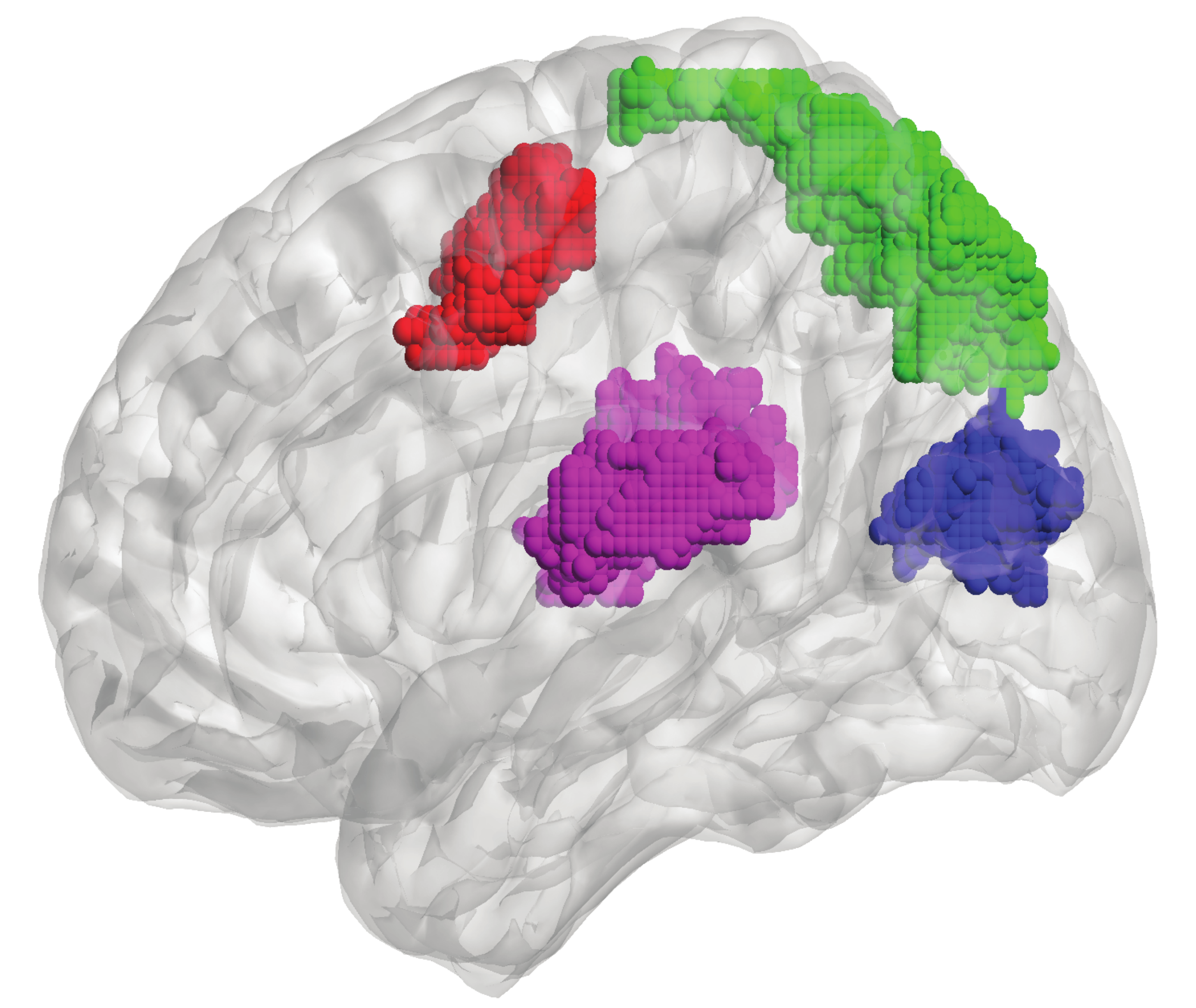}
\caption{
\label{SI::fig:fourModules} Spatial location of four modules: the anterior cingulate AC (red), posterior parietal cortex PPC (green), posterior occipital visual areas V1/V2 (blue), and auditory cortex (magenta) for a typical subject. The three modules, AC, PPC, and V1/V2 appear consistently for all 16 subjects whereas the auditory cortex appears in only 7 out of all 16 participants.}
\end{figure}
\end{center}

Note that the auditory cortex was activated as a major cluster in only 7 out of all 16 subjects in our percolation analysis. 
Fig.\,\ref{SI::fig:fourModules} shows the spatial location of a subject in which the auditory cortex was activated as well.
While a more complete study would also include this cluster, we focused on the brain NoN composed of AC-PPC-V1/V2, which consistently appears for all 16 participants.

\subsection*{Detecting the modules of the brain NoN}

 To detect the modules in the brain NoN we first
  calculate the cross-correlation $C_{ij}$ between the phases of BOLD
  response for each pair of voxels $i$ and $j$: \beq
  C_{ij}\ =\ \frac{1}{N}\sum_{t=1}^N\cos(\phi_i^t-\phi_j^t)\ .
\label{eq:Cij}
\eeq where $N=40$ is the number of measurements of the phases, ie, the
total number that the stimulus is presented to each subject.
The cross-correlation $C_{ij}$ ranges from $-1$ to $1$. $C_{ij}>0$ corresponds 
to positive correlations, $C_{ij}<0$ corresponds negative correlations, and $C_{ij}=0$ 
indicates the lack of correlation between a pair of voxels, $i$ and $j$.

 Then we use a procedure inspired by bond percolation
  \cite{gallos,saulo,eguiluz} to separate the modules, which is
  described next.  We progressively consider the voxels that are
  strongly correlated, and, by using a threshold $T$, we create a
  fictitious link between two voxels $i$ and $j$ if $C_{ij}>T$.
At a certain percolation threshold $T_c$ a largest connected component
emerges, which gradually increases with increasing the fraction of
occupied bond.  Due to the modular structure of the brain, the size of
the largest component increases with a series of jumps when the
threshold $T$ decreases.  This growth pattern of the largest component
in brain reveals that modules defined by strongly correlated
connections merge one by one as $T$ is lowered.  From this
observation, we can naturally identify modules in brain networks
resulting from strong correlations $C_{ij}>T$
\cite{bullmore,gallos,saulo,eguiluz}.  
Notice that we use this
procedure only to identify which voxel belongs to which module, but we
do not use the fictitious links as representative of the intra-links
and inter-links. Therefore, from now on we forget about the fictitious
links and we proceed by inferring the connections between voxels using
a machine learning method, as explained in the next section.

\subsection*{Inferring the connections}

To define the 3NoN composed of AC-PPC-V1/V2 depicted
  in Fig.\,3a we reconstruct the network's intra-links and
  inter-links by using a Machine Learning technique called Maximum
  Entropy Modelling (MEM). The method has been applied to neuronal
  populations in \cite{bialek} and it is similar to methods to infer
  the weights of the paths connecting two brain areas in the
  computational neuroscience community
  \cite{sarkar,robinson2}.  The weight of the links
  that we infer are analogous to what is called direct effective
  connection matrix (deCM) in \cite{robinson2}: they embody the
  strength of each direct connection between points in a given brain
  state. 

This method receives in input the set of
  cross-correlations $\{C_{ij}\}$ of the fMRI signals between pair of
  voxels measured from the fMRI BOLD response in the 3NoN, and outputs
  the intramodular and intermodular weights $\{J_{ij}\}$ of the path
  between $i$ and $j$, also called interaction strengths or couplings
  in statistical physics. A value $J_{ij}\neq 0$ means that there
  exists a link between the pair of voxels $i$ and $j$ and the weight
  of this link is given by the value of $J_{ij}$, while if $J_{ij}= 0$
  then there is no direct connection between $i$ and $j$.

 In order to implement the MEM, we first calculate the
  cross-correlation $C_{ij}$ between the phases of BOLD response for
  each pair of voxels $i$ and $j$ as in Eq.\,(\ref{eq:Cij}).
The cross-correlation $C_{ij}$ ranges from -1 to 1.  $C_{ij}>0$
corresponds to positive correlations, $C_{ij}<0$ corresponds to
negative correlations, and $C_{ij}=0$ indicates the lack of
correlation between a pair of voxels, $i$ and $j$.

 The MEM is based on the the Maximum Entropy
  Principle, which implies that the most general joint distribution
  $P(\phi_1,\dots, \phi_N|\hat{J})$ of the phases $\phi_i\in[0,2\pi]$,
  assuming solely the knowledge of the cross-correlations $C_{ij}$,
  contains only pairwise (i.e. two body) interactions (or equivalently
  weights) $J_{ij}$, and is explicitly given by the following
  expression: \beq P(\phi_1,\dots,
  \phi_N|\hat{J})\ =\ \frac{1}{Z(\hat{J})}\ \prod_{i<j}e^{J_{ij}\cos(\phi_i-\phi_j)}\ .
\label{eq:boltzmann}
\eeq
The goal of this method is to estimate the interactions $\{J_{ij}\}$ such that 
the cross-correlations computed with the measure in Eq.\,(\ref{eq:boltzmann})
match the observed quantities $C_{ij}$, i.e.:
\beq
\la\cos(\phi_i-\phi_j)\ra \equiv \int d\vec{\phi}\ P(\phi_1,\dots, \phi_N|\hat{J})
\cos(\phi_i-\phi_j) = C_{ij}\ .
\eeq
The problem of inferring the interaction matrix $\hat{J}$ from the cross-correlation 
matrix $\hat{C}$ is solved by maximizing the 
log-likelihood $\mathcal{L}(\hat{J}|\hat{C})$:
\beq
\mathcal{L}(\hat{J}|\hat{C})\ =\ \sum_{i<j}J_{ij}C_{ij}-\log Z(\hat{J})\ ,
\label{eq:likelihood}
\eeq
from which the inferred $\hat{J}^*$ is obtained as:
\beq
\hat{J}^*\ =\ {\rm argmax}_{\hat{J}}\ \mathcal{L}(\hat{J}|\hat{C})\ .
\eeq
Indeed, by extremizing $\mathcal{L}(\hat{J}|\hat{C})$ with 
respect to $J_{ij}$ we find 
\begin{equation}
\begin{aligned}
0=\frac{\partial}{\partial J_{ij}}\mathcal{L}&(\hat{J}|\hat{C}) = C_{ij}-\la\cos(\phi_i-\phi_j)\ra \\ 
\rightarrow&\  C_{ij}=\la\cos(\phi_i-\phi_j)\ra.
\end{aligned}
\end{equation}
 


The main difficulty of this method is to compute the quantity $\log
Z(\hat{J})$, the negative of which is called free energy in
statistical physics.  Unfortunately there is no known closed-form for
$\log Z(\hat{J})$, and, as a consequence, also to estimate the
interactions $J_{ij}$ that maximize the log-likelihood
Eq.\,(\ref{eq:likelihood}).

Therefore, to solve the problem, we use a Montecarlo sampling method
to compute the averages $\la\cos(\phi_i-\phi_j)\ra$, and then we use
an approximate iterative gradient ascent algorithm to update the
current estimate of the couplings $J_{ij}$. In practice, we start from
an initial guess $\{J_{ij}^0\}$ at the initial time $t=0$ of the
machine learning algorithm, and then we update the $J_{ij}$'s using
the following rule: \beq J_{ij}^{t+1} = J_{ij}^t - \eta\left[
  \la\cos(\phi_i-\phi_j)\ra^t-C_{ij}\right] +
\alpha(J_{ij}^t-J_{ij}^{t-1}) \ , \eeq where the quantities
$\la\cos(\phi_i-\phi_j)\ra^t$ are the cross-correlations computed via
Montecarlo sampling using the current estimate of the couplings
$J_{ij}^t$ at time $t$; $\eta$ is the learning rate, and $\alpha$ is a
damping factor that we use to help the convergence. We chose the
initial $\{J_{ij}^0\}$ all equal to $0.1$, the learning rate $\eta
=0.01$ and the damping factor $\alpha=0.7$.  

After estimating the couplings $J_{ij}$ we build the 3NoN in two
steps. First of all we establish the intra-links between nodes
(i.e. voxels) belonging to the same module, separately for each
module, and then we connect the nodes in different modules through the
inter-links.  Ideally we would like to put a link between two nodes
$i$ and $j$ if and only if the corresponding $J_{ij}$ is different
from zero.  However, the inference of the couplings $J_{ij}$ is
affected by noise (both because of the uncertainties in the
measurements of the $C_{ij}$ and in the Montecarlo sampling), and thus
we do not have a sharp classification of zero and non-zero
couplings. Therefore, we define the connections by thresholding the
$J_{ij}$ with the following criterion.  First we compute the standard
scores $Z_{ij}$ of the raw couplings $J_{ij}$, defined as $Z_{ij} =
(J_{ij}-\la J\ra)/\sigma$, where $\la J\ra$ and $\sigma$ are the mean
and the standard deviation of the pool $\{J_{ij}\}$.  Then, for each
module separately, we consider a threshold $T$, and we create an
intra-link between two nodes in the same module if $Z_{ij}>T$.

The question of what threshold value $T$ precisely defines the three
networks is resolved using the following procedure. 
First we add intra-links independently in each module  by choosing $T$ 
to be such that the average degree $\langle k^{\rm in}\rangle$ of 
 intra-links is the same for each module, and equal to 
$\langle k^{\rm in}\rangle=5$.

Once the intra-links have been established, we proceed to add inter-links 
between pairs of voxels in different modules. Again, we consider a threshold 
$T$ and we create an inter-link between two nodes $i$ and $j$ in two different 
modules if $Z_{ij}>T$. The threshold $T$ is chosen to be such that the average
$\langle k^{\rm out}\rangle$ of the degree of the  inter-links is 
$\langle k^{\rm out}\rangle=0.5$.




From this procedure we identify three predominant clusters emerging in
all subjects as in previous work of dual-task data \cite{gallos}:
anterior cingulate (AC), posterior parietal cortex (PPC), and
posterior occipital cortex (V1/V2) (Fig.\,3a). 
The average in-degree is $\langle k^{\rm in}\rangle = 5$ and
out-degree $\langle k^{\rm out}\rangle = 0.5$.
The network data for the subject
shown in Fig.\,3 can be downloaded at:
\url{http://www-levich.engr.ccny.cuny.edu/webpage/hmakse/software-and-data}.


\section*{Collective Influence Map of the brain: CI-map }
\label{si-map}

Once we construct the brain NoN, we can directly identify the location
of influential nodes, through the collective influence theory.  First,
we compute the Collective Influence Eq.\,(8) in the
main text for the brain NoN of each subject using $\ell=3$ For other
$\ell$, we found no relevant change of the results, and increasing
$\ell$ leads to degrading the algorithm since the networks are small
and the maximum diameter is reached.  We apply the adaptive CI
algorithm explained in SI Text.
Then, we are able to find the core nodes in the brain for a given
subject according to the CI score. The typical result for the mutually
connected giant component is shown in Fig.\,3 for a given
subject. We identify the most influential nodes in the brain network
as those obtained before the optimal percolation transition at the
critical point $q_{\rm infl}$.  After finding the top CI voxels for
each subject, we obtain the Collective Influence CI-map of the brain
showing the spatial distribution of influencers, averaged over 16
subjects.

Since the number of top influencers (those included up to $q_{\rm
  infl}$) varies with each subject (the number of nodes in the 3NoN is
not the same across subjects), and to facilitate averaging across
different subjects, we measure the ranking of the CI for each voxel
and introduce the normalized influence by following,
\begin{equation}
R_{CI}(i)=\frac{r_0-r_i-1}{r_0},
\end{equation}
where $r_i$ is the ranking of a node $i$ and $r_0$ is the ranking of a
baseline chosen arbitrarily.  $R_{CI}(i)=1$ corresponds to the highest
CI node and $R_{CI}(i)$ decreases with decreasing $r_i$.  In this
study, we set $r_0$ as the ranking of top $15 \%$ node.  Then, we
regard the sum of $R_{CI}(i)$ as the representative influence of a
voxel $i$, over all subjects. In our experiments, the sum of $R_{CI}$
ranges from 0 to $5.2$ 
and the higher value, the more influential region.

The CI-map in Fig.\,3 
reveals the most influential regions in the brain during dual-task
experiments. The spatial distribution of core regions predicted by CI
algorithm is consistent with well-known functions of each modules as
well. To be specific, the most influential regions (top CI nodes) are
mainly located in the AC module which is recruited for top-down and
bottom-up control.  The PPC region contains a smaller portion of
influential nodes next to the AC module since the PPC is responsible
for both top-down and bottom-up control as well.  In contrary, the
influential voxels are rarely located in the V1/V2 module, which is
involved in mostly processing of visual signal and bottom-up control.
We conclude by saying that our theory has recently been tested in rats using pharmacogenetic
interventions targeting the neural influencers responsible for memory consolidation \cite{Min2017}.

\end{document}